\documentclass[apj]{emulateapj}
\usepackage{graphicx,natbib,amsfonts,amssymb,rotating,lscape}
\slugcomment{Submitted to the Astrophysical Journal on January 18, 2006; Accepted July 10, 2006}
\newcommand{\about}     {\hbox{$\sim$}}
\newcommand{\si}[1]{\ensuremath{_{\textrm{\scriptsize{#1}}}}}

\begin{document}

\title{The X-ray Properties of Active Galactic Nuclei with Double-Peaked Balmer Lines}

\author{
Iskra V. Strateva\altaffilmark{1,2},
W. N. Brandt\altaffilmark{1},
Michael Eracleous\altaffilmark{1},
Donald P. Schneider\altaffilmark{1},
George Chartas\altaffilmark{1}
}

\altaffiltext{1}{Department of Astronomy and Astrophysics, 525 Davey Lab, 
Pennsylvania State University, University Park, PA 16802} 
\altaffiltext{2}{Max-Planck-Institut f\"{u}r extraterrestrische Physik, Postfach 1312,
 85741 Garching, Germany}

\addtocounter{footnote}{2} 

\begin{abstract}
  Double-peaked Balmer-line profiles originate in the accretion disks
  of a few percent of optically selected AGN. The reasons behind the
  strong low-ionization line emission from the accretion disks of
  these objects is still uncertain.  In this paper, we characterize
  the X-ray properties of 39 double-peaked Balmer line AGN, 29 from
  the Sloan Digital Sky Survey and 10 low optical-luminosity
  double-peaked emitters from earlier radio-selected samples. We find
  that the UV-to-X-ray slope of radio-quiet (RQ) double-peaked
  emitters as a class does not differ substantially from that of
  normal RQ AGN with similar UV monochromatic luminosity. The
  radio-loud (RL) double-peaked emitters, with the exception of LINER
  galaxies, are more luminous in the X-rays than RQ AGN, as has been
  observed for other RL AGN with single-peaked profiles.  The X-ray
  spectral shapes of double-peaked emitters, measured by their
  hardness ratios or power-law photon indices, are also largely
  consistent with those of normal AGN of similar radio-loudness. In
  practically all cases studied here, external illumination of the
  accretion disk is necessary to produce the Balmer-line emission, as
  the gravitational energy released locally in the disk by viscous
  stresses is insufficient to produce lines of the observed strength.
  In the Appendix we study the variability of Mrk\,926, a
  double-peaked emitter with several observations in the optical and
  X-ray bands.
\end{abstract}    


\section{INTRODUCTION}
\label{intro}

A small class of Active Galactic Nuclei (AGN) emit broad,
double-peaked, low-ionization lines, which most likely originate
directly from the AGN accretion disk. The frequency of double-peaked
lines is high (\about 20\%) among the radio-loud equivalents of
Seyfert galaxies, broad-line radio galaxies
\citep[BLRG;][]{EH94,EH03}, and much lower (\about 3\%) among the
general population of low-redshift, optically selected AGN
\citep[][ hereafter S03]{sdssSample}. The class of double-peaked
emitters is diverse --- it includes both radio-loud (RL) and
radio-quiet (RQ) AGN with luminosities ranging from low-luminosity
low-ionization nuclear emission-line region (LINER) galaxies to
low-luminosity quasars. The optical spectra of double-peaked emitters
are sometimes dominated by the featureless continuum of the AGN, but
in a significant fraction of cases starlight from the host galaxy
makes a dominant contribution to the observed optical continuum. The
handful of double-peaked emission-line BLRG with reliable black-hole
mass measurements \citep[10 black hole masses measured via the
\ion{Ca}{2} triplet velocity dispersion and the relation between
stellar velocity dispersion and black-hole mass;][]{2peakMbh} have
$0.4\times10^8M_{\odot}<M\si{BH}<5\times10^8M_{\odot}$ and range from
low-to-moderate Eddington-ratio AGN
($10^{-5}<L\si{bol}/L\si{Edd}<10^{-1}$). 

\begin{figure}
\epsscale{1.0}
\plotone{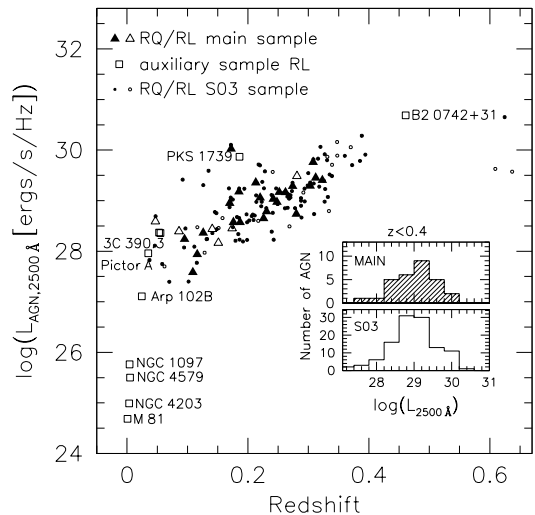}
\caption{Redshift vs. 2500\,\AA\ monochromatic luminosity (corrected
  for starlight). The main SDSS double-peaked sample is shown with
  triangles, the S03 sample with small circles, and the auxiliary
  sample with open squares. Open symbols denote the RL AGN. The
  $z<0.4$ monochromatic-luminosity histograms for the main sample
  (hatched histogram) and the S03 (open histogram) samples are given
  as an inset, showing no significant difference in the two
  distributions.
\label{luvz}} 
\end{figure}

A large fraction of the originally studied double-peaked emitters were
associated with low optical luminosity AGN. The Sloan Digital Sky
Survey \citep[SDSS,][]{York00} sample, which increased the number of
medium-luminosity AGN (see Figure~\ref{luvz}), also showed that the
optical luminosity distributions of double-peaked emitters and normal
AGN at $z\lesssim0.3$ are not substantially different. This suggests that the
apparent lack of double-peaked emitters among the most optically
luminous AGN is a selection effect related to the rarity of very
luminous sources and the small volume of space considered to date. The
large range of Eddington ratios observed in AGN with double-peaked
emission lines together with their large range of luminosities (which
probably translates into a large range of Eddington ratios) suggest
that the existence of strong Balmer-line emission from the accretion
disks of AGN is either independent of the Eddington ratio of the
source, or that there are different explanations for the presence of
double-peaked Balmer lines in AGN with different Eddington ratios.  In
addition to a large range of Eddington ratios, the known double-peaked
emitters tend to have broader Balmer lines ($\approx5\times$ the full
width at half maximum of normal AGN) and stronger low-ionization
narrow lines (e.g., [\ion{O}{1}]~6300\AA) than similar AGN without
double-peaked lines. As a class, the double-peaked emitters possess a
diverse set of optical/UV properties but most do not appear remarkably
different in their other properties from the general AGN population.

Based on the prototype double-peaked emitter, Arp\,102B, and the
initial sample of double-peaked BLRG and LINER galaxies, there were
reasons to believe that disk emission in AGN is associated with
low-luminosity, low Eddington-ratio accretion in which the inner parts
of the accretion disk (less than a few tens of gravitational radii,
$R_g$) became geometrically thick and optically thin
\citep[e.g.,][]{adaf}, emitting energetic X-rays which illuminate the
outer (outside a few hundred gravitational radii) regions of the disk
\citep{CH89,ADAF2peaked,EH03}. Since most of the initially known
double-peaked emitters had strong Balmer lines, which could not have
been produced by the release of gravitational energy locally without
invoking an unrealistic radiative efficiency, the X-ray illumination
from the thick inner disk was necessary to produce strong
disk-emission signatures.

The theoretical model presented above remains the best explanation for
low-ionization broad-line disk emission in low-luminosity, low
Eddington-ratio AGN. In its original form, however, it is not
applicable to higher Eddington-ratio (and luminosity) double-peaked
emitters. As the Eddington ratio and luminosity of the AGN increases,
the structure of the inner disk and the illumination pattern of the
outer disk change. In addition, line-driven winds could become
important, filling-in the low-velocity minima of the double-peaked
emission lines until the profiles are effectively single peaked
\citep[e.g.,][]{winds2,winds}. If we can establish clearly the need for
external illumination in higher luminosity (and presumably
Eddington ratio) sources, we can revise the theoretical picture of
their central regions by exploring evidence for the presence of a
different kind of disk-illuminating structure --- for example --- a
``flared'' disk, a scattering corona, vertically extended wind, or a
radiation supported torus.

In search of clues that might reveal the mechanism responsible for the
existence of double-peaked Balmer lines in AGN, we decided to study in
more detail the 0.5--10\,keV X-ray properties of this class. If
additional illumination is necessary to produce significant disk
line-emission signatures in AGN, their high-energy emission and their
overall spectral energy distributions might be expected to differ from
those of single-peaked AGN. S03 and \citet{Wu04} calculated the
UV-to-X-ray slopes ($\alpha$\si{ox}) of double-peaked emitters with
\emph{ROSAT} All Sky Survey \citep[RASS,][]{rass} detections and
concluded that they do not differ strongly from those of other
AGN. These rough comparisons, however, included only X-ray detected
subsamples of double-peaked emitters, which are not representative of
the full double-peaked Sloan Digital Sky Survey sample and the general
double-peaked line population. They also did not take
into account the dependence of $\alpha$\si{ox} on AGN luminosity
\citep[e.g.,][]{aox05,Steffen05}.

In this paper we present a detailed investigation of the X-ray
properties (soft X-ray luminosities, UV-to-X-ray slopes, X-ray
power-law photon indices, and X-ray absorption whenever available) of
a sample of double-peaked emitters serendipitously included in pointed
\emph{ROSAT}, \emph{XMM-Newton}, and \emph{Chandra} observations. The
paper is organized as follows: the sample of double-peaked emitters
with X-ray observations is presented in \S~\ref{sample}, followed by
the study of their UV-to-X-ray slopes and X-ray spectral shapes in
\S~\ref{sec3}. \S~\ref{ebudget} discusses the evidence for external
illumination, followed by the summary and conclusions in
\S~\ref{sec4}. In the Appendix we comment on the variability of
Mrk\,926, a double-peaked emitter from our sample with several optical
and X-ray observations. Throughout this work we use the
\emph{Wilkinson Microwave Anisotropy Probe} cosmology parameters from
\citet{Spergel03} to compute the luminosities of AGN:
$\Omega_{\Lambda}=0.73$, flat cosmology, with
$H_0$=72\,km\,s$^{-1}$\,Mpc$^{-1}$.

\section{SAMPLE SELECTION}
\label{sample}

Our sample consists of 29 double-peaked emitters selected from the
SDSS and serendipitously observed or previously targeted by
\emph{ROSAT}, \emph{XMM-Newton}, or \emph{Chandra} (hereafter the main
sample). We add to this sample 10 BLRG, LINER galaxies, and RL quasars
with good UV and X-ray observations (hereafter the auxiliary sample).
A total of 16 of the 39 double-peaked emitters are RL (6 main-sample
objects and 10 auxiliary sample objects), and the remaining 23
(main-sample objects) are RQ.

\subsection{The Main Sample of Double-Peaked Emitters}
\label{mainS}

In order to increase the number of SDSS double-peaked emitters with
sensitive X-ray observations, we start with a sample of SDSS
AGN\footnote{Here the term AGN is not restricted to represent the SDSS
quasar sample presented by \citet{dr3qso}; it stands for all objects
showing broad Balmer lines, including those originally targeted as
galaxies by the SDSS.} with $z<0.4$ (which guarantees that the
H$\alpha$ line is observed) selected from Data Release 3
\citep[DR3,][]{DR3} which fall in pointed \emph{ROSAT},
\emph{XMM-Newton}, or \emph{Chandra} observations that are publicly
available. The effective exposure times (i.e., after accounting for
the off-axis angle, X-ray background flares, etc.) of these
observations are all $>1$\,ks and are $>2$\,ks for 80\% of the
objects; the median exposure time is 4.2\,ks. Using SDSS-RASS matches
would have substantially increased our sample size but decreased our
detection fraction and the total X-ray counts of detected objects
(limiting our ability to extract X-ray spectral information). Using
only the S03 double-peaked emitter sample, which was selected from SDSS
observations performed prior to March 2003, would have decreased the
number of sources, thus weakening the significance of our statistical
inferences.

Since the double-peaked emitters are a subsample of the asymmetric
broad-line AGN (see S03 for more details), we start by selecting a
subsample of the SDSS AGN with \emph{ROSAT}, \emph{XMM-Newton}, or
\emph{Chandra} observations and asymmetric broad H$\alpha$ emission
lines. We fit the H$\alpha$ line region of each spectrum\footnote{The
spectra were decomposed into AGN and starlight components first
\citep{galfits}, and both the starlight and a power-law representation
of the AGN continuum were subtracted.}  with a combination of
Gaussians.  These Gaussian representations of the H$\alpha$ line
region allow us to isolate the broad H$\alpha$ line component
(excluding the narrow H$\alpha$, [\ion{N}{2}], and [\ion{S}{2}] lines)
and to select the AGN with double-peaked broad lines. In addition,
obtaining a smooth representation of the broad H$\alpha$ line
component allows easy measurement of a set of model-independent
parameters: (1) the rest-frame full width at half maximum, FWHM, (2)
the rest-frame full width at quarter maximum, FWQM, (3,4) the
centroids of the line at half and quarter maximum, FWHMc and FWQMc
(measured with respect to the rest-frame narrow H$\alpha$ line), (5,6)
the rest wavelengths of the red and blue peaks with respect to the
narrow H$\alpha$ line, $\lambda$\si{Red} and $\lambda$\si{Blue}, and
(7) the peak separation, $\lambda\si{Red}-\lambda\si{Blue}$.  More
details on the broad-line parameter measurements and the double-peak
selection criteria can be found in \S3.2 of S03; the H$\alpha$ line
measurements for the selected objects are reported in
Table~\ref{tab7}. Eight of the resulting 29 main-sample double-peaked
emitters are part of the S03 sample. For consistency with the rest of
the sample, we used the newest SDSS spectroscopic reductions currently
available and obtained new H$\alpha$-line parameter measurements for
these eight objects. The results are consistent with those reported in
the S03 paper, within the errors quoted in Table~4 of S03.  The
H$\alpha$-line luminosities of the eight new reductions were found to
be slightly higher on the average (by \about 20\%).

\begin{deluxetable*}{ccccrrrrrr} 
\tablewidth{0pt} 
\tablenum{1}
\tablecaption{Main Sample: H$\alpha$-Line Measurements and \hbox{0.5--2\,keV} Luminosities} 
\tablehead{\colhead{Object} & \colhead{$z$} & \colhead{$L\si{H$\alpha$}$} & \colhead{$L\si{0.5--2\,keV}$} & \colhead{FWQM} & \colhead{FWQMc} & \colhead{FWHM} & \colhead{FWHMc}  & \colhead{$\lambda\si{Red}$} & \colhead{$\lambda\si{Blue}$} \\
\colhead{(1)} & \colhead{(2)} & \colhead{(3)} & \colhead{(4)} & \colhead{(5)} & \colhead{(6)} & \colhead{(7)} & \colhead{(8)} & \colhead{(9)} & \colhead{(10)}}
\startdata
004319.75$+$005115.3 & 0.3081 &  17  & 180 & 14600 &      10 & 11900 &  $-$210 &  3200 & $-$3400 \\
005709.93$+$144610.3 & 0.1722 &  35  & 240 & 12900 &    1230 &  9570 &    1010 &  2300 & $-$1010 \\
022417.17$-$092549.3  & 0.3115 &  12  & 84   &  9800 &      90 &  7160 &     270 &  1860 &  $-$240 \\
081329.29$+$483427.9 & 0.2737 &  10  & 90   & 11800 &  $-$190 &  8610 &  $-$530 &  1350 & $-$1970 \\
091828.60$+$513932.1 & 0.1854 &  6.2 & 8.2  &  6350 &   $-$20 &  4820 &   $-$70 &   560 &  $-$940 \\
092108.63$+$453857.4 & 0.1744 &  1.5 & 87   & 17120 & $-$3520 &  9270 & $-$1320 &   130 & $-$3590 \\
092813.25$+$052622.5 & 0.1874 &  4.3 & 18   & 13800 &     690 &  9660 &     920 &  3180 &  1200 \\
093844.46$+$005715.8 & 0.1704 &  31  & 37  & 11900 &     230 &  8250 &   $-$40 &  2820 &  $-$580 \\
094215.13$+$090015.8 & 0.2126 &  11  & 18   & 42700 &    4380 & 36900 &    3020 & 20100 & $-$12400 \\
094745.15$+$072520.6 & 0.0858 &  5.2 & 3.1  & 15400 &  $-$980 &  9190 &  $-$240 & $-$770 & $-$6630 \\
095427.61$+$485638.1 & 0.2481 &  4.5 & 25   & 15100 &     380 &  8520 &  $-$500 &  6150 & $-$2460 \\
095802.84$+$490311.1 & 0.2416 &  4.9 & 22   &  7800 &      90 &  5760 &  $-$100 &   640 & $-$1000 \\
104132.78$-$005057.5  & 0.3029 &  9.6 & 120 & 10600 &    1000 &  7340 &    1180 &  1020 & $-$1290 \\
110109.58$+$512207.1 & 0.2521 &  4.2 & 22   & 12000 &     420 &  9260 &     420 &  1960 &  $-$620 \\
111121.71$+$482046.0 & 0.2809 &  10  & 31   & 23800 &     710 & 18800 &     260 &  6670 & $-$3980 \\
113450.99$+$491208.9 & 0.1765 &  1.6 & 2.6  &  9850 &  $-$190 &  6530 &  $-$280 & $-$110 & $-$1960 \\
114454.86$+$560238.2 & 0.2310 &  3.0 & 29   & 10900 &     970 &  8310 &     670 &  3470 & $-$1000 \\
114719.22$+$003351.2 & 0.2624 &  7.1 & $<$10 &  9170 &  $-$160 &  6830 &  $-$160 &  1500 & $-$1380 \\
115038.86$+$020854.2 & 0.1095 & 0.41& 3.1  &  8950 &     270 &  7080 &   $-$40 &   320 & $-$2100 \\
115741.94$-$032106.1  & 0.2195 &  3.1 & $<$7.3 & 11600 &  $-$170 &  8890 &   $-$50 &  1920 & $-$1670 \\
120848.82$+$101342.6 & 0.1158 & 0.62& 21   &  9450 &   $-$10 &  7350 &     330 &  2410 &  $-$150 \\
130723.13$+$532318.9 & 0.3231 &  8.0 & 110 & 10200 &  $-$120 &  7990 &  $-$170 &  1950 & $-$1950 \\
132355.69$+$652233.0 & 0.2261 &  1.6 & 5.8  &  7610 &     100 &  5720 &     460 &  2910 &   $-$40 \\
144302.76$+$520137.2 & 0.1412 &  3.2 & 73   &  9170 &   $-$70 &  5830 &  $-$470 &  3280 &  $-$520 \\
161742.53$+$322234.3 & 0.1500 &  12  & 6.2  & 23300 &    1150 & 19600 &     840 &  7400 & $-$5490 \\
164031.86$+$373437.2 & 0.2796 &  3.1 & 56   &  6790 &  $-$400 &  4970 &  $-$800 & $-$10 & $-$2030 \\
170102.29$+$340400.6 & 0.0946 &  3.0 & 8.0  &  7910 &  $-$260 &  5800 &  $-$450 & $-$90 & $-$1760 \\
213338.42$+$101923.6 & 0.1257 &  1.4 & $<$2.2 & 10400 &     280 &  7970 &      30 &  2270 & $-$1980 \\
230443.47$-$084108.6  & 0.0469 &  4.8 & 85  & 11400 & $-$1210 &  8790 &  $-$970 &  1560 & $-$1620 \\
\enddata
\tablecomments{(1) the SDSS name given in the J2000 epoch RA and Dec
  form, HHMMSS.ss$\pm$DDMMSS.s; (2) redshift; (3) H$\alpha$ line
  luminosity in units of $10^{42}$\,erg\,s$^{-1}$; (4) rest-frame
  0.5--2\,keV luminosity in units of $10^{42}$\,erg\,s$^{-1}$,
  estimated from the observed 0.5--2\,keV band for the 22 \emph{ROSAT}
  objects (assuming $\Gamma=2$) and estimated from the spectral fits
  for the 7 \emph{XMM-Newton} and \emph{Chandra} observed objects,
  including corrections for any intrinsic absorption; (5) the FWQM of the
  H$\alpha$ line, in km\,s$^{-1}$; (6) the FWQM centroid in
  km\,s$^{-1}$; (7) the FWHM of the H$\alpha$ line, in km\,s$^{-1}$;
  (8) the FWHM centroid in km\,s$^{-1}$; (9) the position of the red
  peak, $\lambda\si{Red}$, with respect to the narrow H$\alpha$
  line, in km\,s$^{-1}$; (10) the position of the blue peak,
  $\lambda\si{Blue}$, with respect to the narrow H$\alpha$
  line, in km\,s$^{-1}$. Positive velocities denote a redshift.}
\label{tab7}
\end{deluxetable*}

Of the 29 SDSS AGN with double-peaked lines, 22 were observed by the
\emph{ROSAT} PSPC only, five by \emph{XMM-Newton}, and two by
\emph{Chandra}, as shown in Table~\ref{tab1}.  Whenever hard-band
(above 2\,keV) observations were available for double-peaked emitters
in \emph{ROSAT} PSPC fields (3 cases with both \emph{XMM-Newton} and
\emph{ROSAT} data), we used the \emph{XMM-Newton} data.  If both
\emph{XMM-Newton} and \emph{Chandra} data were available, we chose the
observation with the larger number of photons, and compare the results
from the two observatories.  Four SDSS double-peaked emitters have
been observed repeatedly with various X-ray satellites (see the top
portion of Table~\ref{tab1}); we use all the available X-ray data for
the best example, SDSSJ\,2304$-$0841, to study variability in
Appendix~\ref{sec:app}. Five of the X-ray observed double-peaked
emitters have sufficient counts (\hbox{100--24000} in the 2--10\,keV
band) to allow fitting of their X-ray spectra. Three of the objects
observed by \emph{ROSAT} are not detected in the soft
(\hbox{0.5--2\,keV}) band; for these three we derive ($3\sigma$) upper
limits to the soft-band fluxes.

\begin{deluxetable*}{cccccrc} 
\tablewidth{0pt} 
\tablenum{2}
\tablecaption{X-ray Observations} 
\tablehead{\colhead{Object} &\colhead{Observatory} &\colhead{ObsID} &\colhead{Instrument} &\colhead{ObsDate} &\colhead{$t_{\textrm{exp}}$} & \colhead{Comments}\\ \colhead{(1)} &\colhead{(2)} &\colhead{(3)} &\colhead{(4)} &\colhead{(5)} &\colhead{(6)} &\colhead{(7)}}
\startdata
 022417.17$-$092549.3 & \emph{ROSAT} & rp800016n00 & PSPC & 18/01/92 & 13.6 & \\
 081329.29$+$483427.9 & \emph{ROSAT} & rp700249n00 & PSPC & 01/04/91 & 6.5 & \\
 092108.63$+$453857.4 & \emph{ROSAT} & rp700539n00 & PSPC & 03/05/92 & 4.3 & Target \\
 092813.25$+$052622.5 & \emph{ROSAT} & rp200466n00 & PSPC & 12/05/92 & 3.7 & \\
\smallskip
 094215.13$+$090015.8 & \emph{ROSAT} & rp800481n00 & PSPC & 03/11/93 & 1.1 & \\
 094745.15$+$072520.6 & \emph{ROSAT} & rp701587n00 & PSPC & 04/11/93 & 11.0 & Target \\
 095427.61$+$485638.1 & \emph{ROSAT} & rp700046n00 & PSPC & 13/04/91 & 3.5 & \\
 095802.84$+$490311.1 & \emph{ROSAT} & rp700150a02 & PSPC & 20/10/93 & 1.9 & \\
 104132.78$-$005057.5 & \emph{ROSAT} & rp800194n00 & PSPC & 04/06/92 & 7.2 & \\
\smallskip
 110109.58$+$512207.1 & \emph{ROSAT} & rf201357n00 & PSPC & 29/11/92 & 2.1 & \\
 114454.86$+$560238.2 & \emph{ROSAT} & rp800106n00 & PSPC & 19/11/91 & 4.9 & \\
 114719.22$+$003351.2 & \emph{ROSAT} & rp201242n00 & PSPC & 23/06/92 & 5.8 & \\
 115038.86$+$020854.2 & \emph{ROSAT} & rp200813n00 & PSPC & 02/06/92 & 7.7 & \\
 115741.94$-$032106.1 & \emph{ROSAT} & rp201367n00 & PSPC & 07/07/92 & 3.3 & \\
\smallskip
 120848.82$+$101342.6 & \emph{ROSAT} & rp700079a01 & PSPC & 03/06/92 & 3.0 & \\
 130723.13$+$532318.9 & \emph{ROSAT} & rp300394n00 & PSPC & 10/11/93 & 12.0 & \\
 132355.69$+$652233.0 & \emph{ROSAT} & rp700803n00 & PSPC & 30/11/92 & 8.4 & \\
 144302.76$+$520137.2 & \emph{ROSAT} & rp701408n00 & PSPC & 13/07/93 & 6.8 & Target  \\
 161742.53$+$322234.3 & \emph{ROSAT} & rp701589n00 & PSPC & 18/08/93 & 10.1 & 3C\,332, Target\\
\smallskip
 164031.86$+$373437.2 & \emph{ROSAT} & rp800503n00 & PSPC & 31/07/93 & 5.2 & \\
 170102.29$+$340400.6 & \emph{ROSAT} & rp201079n00 & PSPC & 02/09/92 & 6.4 & \\
 213338.42$+$101923.6 & \emph{ROSAT} & rp701252n00 & PSPC & 29/05/93 & 18.6 & \\
\hline
004319.75$+$005115.3 & \emph{XMM-Newton} & 0090070201 & PN/MOS  & 04/01/02 & 16.3 & UM\,269, Target\\ 
...                  & \emph{ASCA}       & 75020000   & GIS/SIS & 13/07/97 & 31.8 & UM\,269, Target\\ 
...                  & \emph{ROSAT}      & rp700377   & PSPC    & 30/12/91 & 5.1  & UM\,269 \\ 
005709.93$+$144610.3 & \emph{Chandra}    & 00865N002  & ACIS-S7 & 28/07/00 & 4.7  & Target, pileed-up \\
091828.60$+$513932.1 & \emph{XMM-Newton} & 0084230601 & PN/MOS  & 26/04/01 & 16.0 & \\ 
...                  & \emph{Chandra}    & 00533N001  & ACIS-I1 & 05/09/00 & 11.3 & \\ 
093844.46$+$005715.8 & \emph{Chandra}    & 04035N001  & ASIS-S2 & 02/01/03 & 1.4 & Target\\ 
111121.71$+$482046.0 & \emph{XMM-Newton} & 0104861001 & PN/MOS  & 01/06/02 & 27.4 & \\
...                  & \emph{XMM-Newton} & 0059750401 & PN/MOS  & 26/04/02 & 29.2$^{*}$ & flaring background \\
...                  & \emph{ROSAT}      & rp700297n00 & PSPC & 15/05/92 & 3.1 & \\
113450.99$+$491208.9 & \emph{XMM-Newton} & 0149900201 & PN/MOS & 24/11/03 & 17.9 & \\ 
230443.47$-$084108.6 & \emph{XMM-Newton} & 0109130701 & PN/MOS & 01/12/00 & 7.3 & MCG-2-58-22, Mrk\,926, Target\\ 
...                  & \emph{ASCA} & 75049010 & GIS/SIS & 15/12/97 & 102.7 & MCG-2-58-22, Mrk\,926, Target \\ 
...                  & \emph{ASCA} & 75049000 & GIS/SIS & 01/06/97 & 104.9 & MCG-2-58-22, Mrk\,926, Target\\ 
...                  & \emph{ASCA} & 70004000 & GIS/SIS & 25/05/93 & 85.6 & MCG-2-58-22, Mrk\,926, Target \\ 
...  & \emph{ROSAT}  & rp701364 & PSPC & 01/12/93 & 2.16 & MCG-2-58-22, Mrk\,926 \\ 
...  & \emph{ROSAT}  & rp700998 & PSPC & 24/05/93 & 10.2 & MCG-2-58-22, Mrk\,926 \\ 
...  & \emph{ROSAT}  & rp701250 & PSPC & 21/05/93 & 18.1 & MCG-2-58-22, Mrk\,926 \\ 
...  & \emph{ROSAT}  & rp700107 & PSPC & 21/11/91 & 3.37 & MCG-2-58-22, Mrk\,926 \\ 
...  & \emph{ROSAT}  & rs931862 & PSPC & 21/11/90 & 0.314 & MCG-2-58-22, Mrk\,926 \\ 
...  & \emph{ROSAT}  & rs931962 & PSPC & 18/11/90 & 0.314 & MCG-2-58-22, Mrk\,926 \\ 
\enddata
\tablecomments{(1) SDSS name given in the J2000 epoch RA and Dec
 form, HHMMSS.ss$\pm$DDMMSS.s; (2) Observatory name; (3) Observation
 ID; (4) Instrument; (5) Date of the observation in a dd/mm/yy format;
 (6) Effective exposure time, $t_{\textrm{exp}}$, in ks, except in
 the case of the second SDSSJ\,111121.71$+$482046.0 \emph{XMM-Newton}
 exposure (denoted by *); this exposure is fully flared and the quoted
 $t_{\textrm{exp}}$ includes the time lost due to flaring; (7)
 Alternative names and comments.}
\label{tab1}
\end{deluxetable*}

Four of the objects observed with \emph{ROSAT} were the targets of
their respective PSPC observations (see the top panel of
Table~\ref{tab1}). Four additional objects, for which X-ray spectra
are available, were also the targets of their respective observations
(see the bottom panel of Table~\ref{tab1}). A large fraction of
targeted observations could have biased the sample of X-ray observed
double-peaked emitters; \S~\ref{representative} below demonstrates
that this is not the case for our sample.

\begin{figure}
\epsscale{1.0}
\plotone{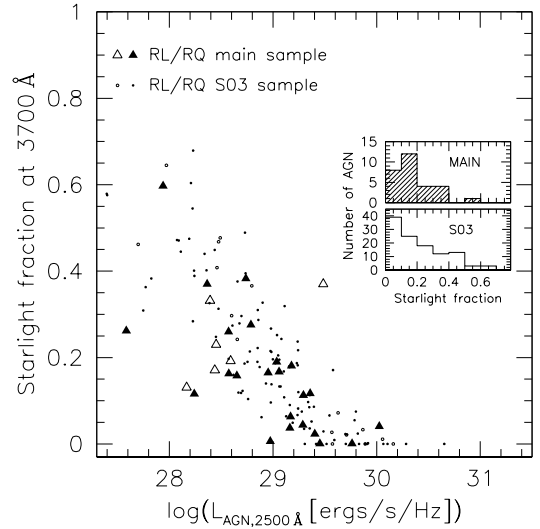}
\caption{Monochromatic luminosity at 2500\,\AA\ vs. starlight fraction.
Symbols are as in Figure~\ref{luvz}. The two histograms show the
distribution of starlight fractions for the main and S03 samples.
\label{gFrac}} 
\end{figure}

\subsection{Main Sample Properties}
\label{representative}

In this section we show the basic properties of the main sample of
double-peaked emitters studied here. We demonstrate that the main
sample is representative of the sample of SDSS double-peaked emitters
by performing a series of one- and two-dimensional statistical
comparisons with the SDSS sample from S03 (hereafter, the S03
sample\footnote{We exclude the three $z\approx0.6$ S03 double-peaked
emitters from the quantitative comparisons presented here, since they
lie outside the standard H$\alpha$ selection region and were included
in the S03 sample after being discovered serendipitously. See S03 for
more details.}). We use Kolmogorov-Smirnov (K-S) tests which measure
the maximum distance, $D$, between two cumulative distributions, and
compute the null hypothesis probability that the data sets are drawn
from the same distribution, $P$. Small values of $D$ and large values
of $P$ indicate that the observed data are highly probable given the
null hypothesis; we will consider this sufficient evidence that the
X-ray observed AGN are representative of the S03 sample. For more
details on the ``two-dimensional K-S test'', see \citet{KS2d}.

\subsubsection{Monochromatic Luminosity and Redshift Distributions}

\begin{figure}
\epsscale{1.0}
\plotone{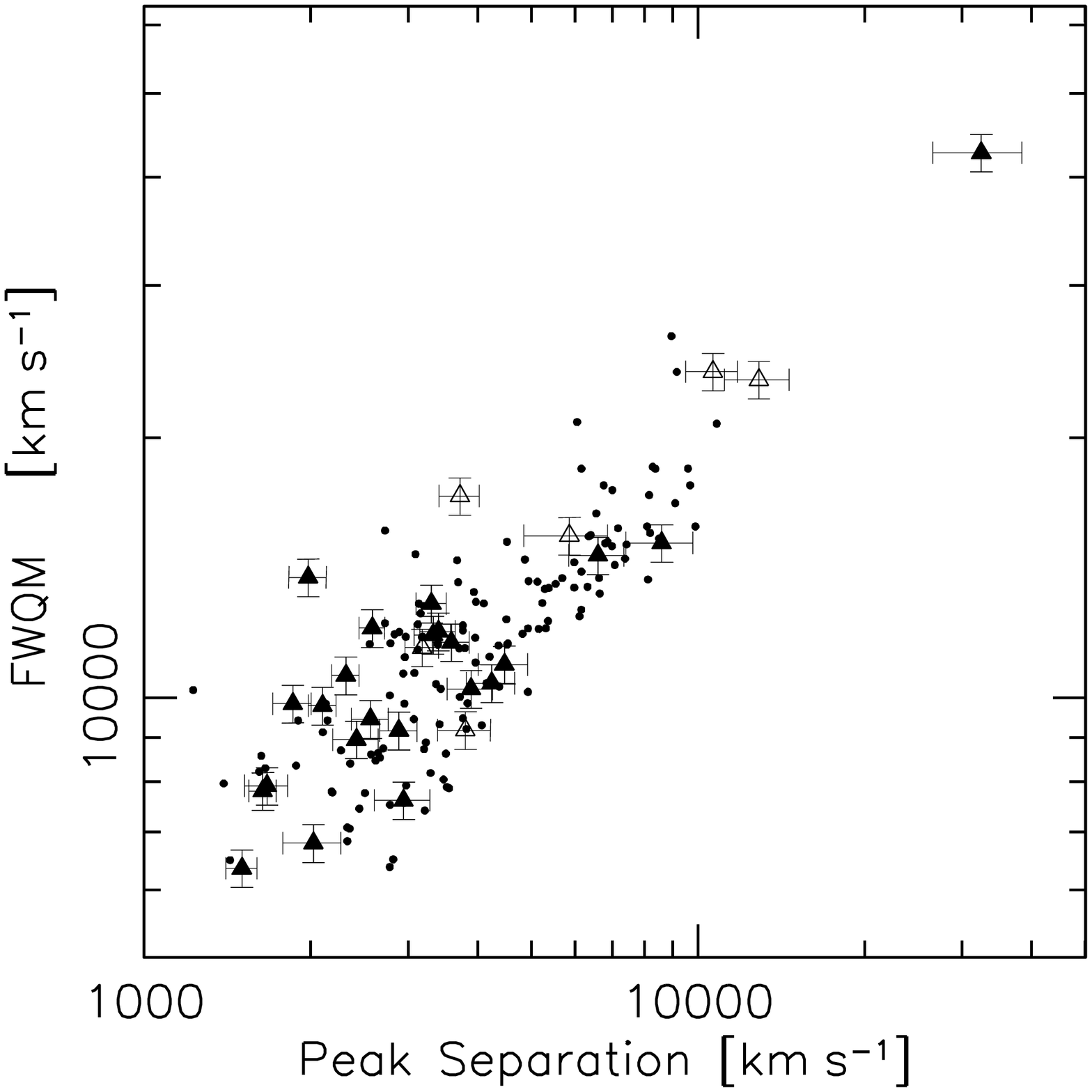}
\caption{Broad H$\alpha$ line parameter comparison: peak separation 
vs. FWQM for the main-sample double-peaked emitters (large open/solid
triangles with error bars for RL/RQ AGN) and the S03 sample (solid
circles).  The error bars (which include uncertainties due to both
measurement error and variability) were estimated to be \about 5\% for
the FWQM measurements and as much as 50\% for the peak-separation
measurements by S03.
\label{PsepFWQM}} 
\end{figure}

Using the starlight-subtracted monochromatic flux measurement at a
rest-frame wavelength of 3700\,\AA\ (the bluest wavelength observed
for all SDSS AGN, irrespective of their redshift), we can estimate the
monochromatic flux at rest-frame 2500\,\AA, which is a convenient
measure of brightness commonly used, for example, to compute the
UV-to-X-ray index. We assumed that the optical/UV flux is a power-law
function of wavelength in the 2500--3700\,\AA\ region ($f_{\lambda}
\propto \lambda^{\alpha_{\lambda}}$) with an index of
\hbox{$\alpha_{\lambda}=-1.5$}
\citep{sdssComposite}. Figure~\ref{luvz} shows the redshift vs. $\log$
of the 2500\,\AA\ monochromatic luminosity
($l\si{2500\AA}=\log[L\si{2500\AA}(\textrm{erg\,cm$^{-2}$\,s$^{-1}$\,Hz$^{-1})$}]$)
diagram for the main SDSS and auxiliary double-peaked samples in
comparison to the S03 sample. The correlation between redshift and
luminosity in flux-limited samples is obvious for both the S03 and the
main samples; in addition, the double-peaked emitters from the main
sample follow the luminosity-redshift trend of the S03 sample. As
shown in the inset histograms in Figure~\ref{luvz}, the monochromatic
luminosities of the main and S03 samples are indistinguishable, which
is confirmed by a one-dimensional K-S test ($D=0.15$, $P=65$\%). The
redshift distribution of the main sample of double-peaked emitters is
also similar to that of the S03 sample ($D=0.17$, $P=47$\%). The
two-dimensional (luminosity-redshift) K-S test gives a similar result,
with $D=0.17$ and $P=57$\%. These results remain qualitatively the
same if we repeat the test for the RQ subsamples.

There are six RL AGN out of a total of 29 AGN (21\%) in the main SDSS
double-peaked sample, compared to 15 out of 113 (13\%) in the
similar-redshift ($z<0.4$) S03 sample. A Fisher exact test confirms
that the difference is not statistically significant, with a chance
probability of 38\%. A careful examination of Figure~\ref{luvz}
reveals that five of the six RL objects in our main sample occupy the
lowest redshift and UV luminosity corner of the diagram. Performing a
two-dimensional K-S test for the RL subsamples of the main and S03
samples returns $D=0.47$ and $P=23$\%. The distance between the RL
distributions is substantial, but not statistically significant.

\subsubsection{Starlight Fraction}

Figure~\ref{gFrac} shows the starlight fractions (measured at
rest-frame 3700\,\AA) of the main-sample double-peaked emitters in
comparison with those of the S03 sample. As a class, RL double-peaked
emitters were found to have larger starlight contributions to the
continuum around H$\alpha$ than other RL AGN. A one-dimensional K-S
test returns a small distance ($D=0.15$) which is not statistically
significant ($P=63$\%). A two-dimensional (starlight
fraction-luminosity) K-S test returns a larger distance ($D=0.28$)
which is also not statistically significant ($P=9$\%).

\subsubsection{Model-Independent Line Measurements}

In addition to luminosity, redshift, and starlight-fraction
distribution comparisons, we also compare the H$\alpha$ line profiles
of the main-sample AGN with those of S03, using the broad-line
parameter measurements described in \S~\ref{mainS}.
Figure~\ref{PsepFWQM} shows the distribution of peak separations
vs. FWQM measurements for the main and S03 samples. As described in \S
6 of S03, the peak separation is largely determined by the outer
radius of the disk emission region for axisymmetric disks, while the
FWQM is determined by the disk inclination and inner radius of
emission. Quantitative comparison of the two samples using a
two-dimensional K-S test shows no evidence that the two samples are
significantly different (with $D=0.18$ and $P=54$\%).

\begin{figure}
\epsscale{1.0}
\plotone{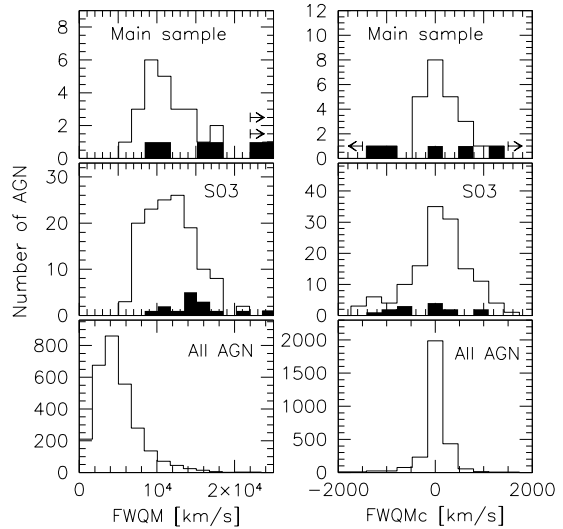}
\caption{Comparison between the distributions of the H$\alpha$ FWQM
(left) and the FWQM centroid (right) for the X-ray observed SDSS AGN
(top histograms), the S03 sample (middle histograms), and the parent
sample of all SDSS AGN with $z\lesssim0.3$ from S03 (bottom
histograms). The solid histograms in the top and middle row give the
contribution of RL double-peaked emitters from the main and S03
samples, respectively; the arrows indicate four main sample
double-peaked emitters with values outside the displayed range.
\label{paramHist}} 
\end{figure}

Figure~\ref{paramHist} shows binned distributions of the FWQM and the
FWQM centroid for the main and S03 samples of double-peaked emitters
(as in Fig.~10 of S03, the histograms of the parent sample of all AGN
with $z\lesssim0.3$ are also shown). In both cases a one-dimensional
K-S test suggests that the (unbinned) distributions of the main and
S03 samples of double-peaked emitters are indistinguishable
($D\approx0.11$, $P\approx90$\%).

\begin{deluxetable*}{lccccccc} 
\tablewidth{0pt} 
\tablenum{3}
\tablecaption{Axisymmetric Disk Fits}
\tablehead{\colhead{Object} &\colhead{$i^{\circ}$} &\colhead{$q$} &\colhead{$\xi_1$} &\colhead{$\xi_2$} &\colhead{$\sigma$ }&\colhead{$F_{\lambda,o}$} &\colhead{$F\si{H$\alpha$}$} \\
\colhead{(1)} &\colhead{(2)} &\colhead{(3)} &\colhead{(4)} &\colhead{(5)} &\colhead{(6)}&\colhead{(7)} &\colhead{(8)} }
\startdata
091828.60$+$513932.1 & 25 & 2 &  100 &  9150 &   495 & 3.0E-4  & 5.5\\
111121.71$+$482046.0 & 36 & 3 &  190 &  2200 &  2000 & 6.5      & 3.2\\ 
113450.99$+$491208.9 & 32 & 2 &  100 &  5530 &   620 & 1.1E-4   & 1.6\\ 
114454.86$+$560238.2 & 16 & 3 &  210 &  2190 &   980 & 3.3      & 1.5\\ 
115741.94$-$032106.1 & 26 & 2 &  230 &  3450 &   900 & 6.7E-3   & 1.9\\ 
130723.13$+$532318.9 & 27 & 2 &  330 &  4260 &   760 & 6.6E-3   & 1.7\\ 
\enddata
\tablecomments{Axisymmetric disk fits following \citet{CH89}. The
inclination (col. 2) is in degrees; the illumination power-law slope,
$q$ (3); the inner (4) and outer (5) radii are in gravitational radii
($R_G$); the turbulent velocity (6) is in km\,s$^{-1}$; the
flux-density normalization (7) is in units of
$10^{-17}$\,erg\,cm$^{-2}$\,s$^{-1}$\,\AA$^{-1}$; the H$\alpha$ line
flux (8) is in units of $10^{-14}$\,erg\,cm$^{-2}$\,s$^{-1}$.}
\label{tab2}
\end{deluxetable*}

\subsubsection{Axisymmetric Model Disk Fits}
\label{fits}

\begin{figure*}
\epsscale{1.0}
\plottwo{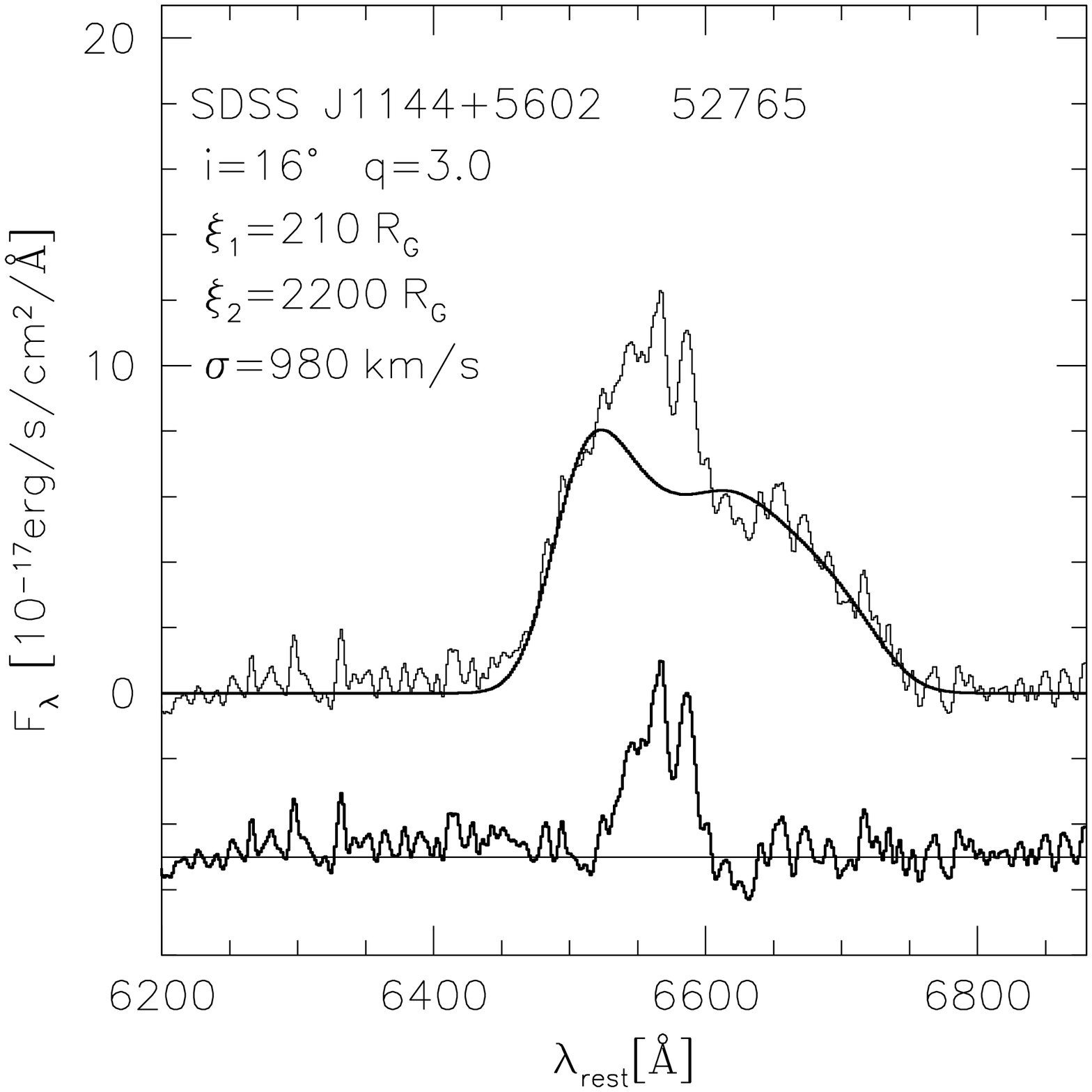}{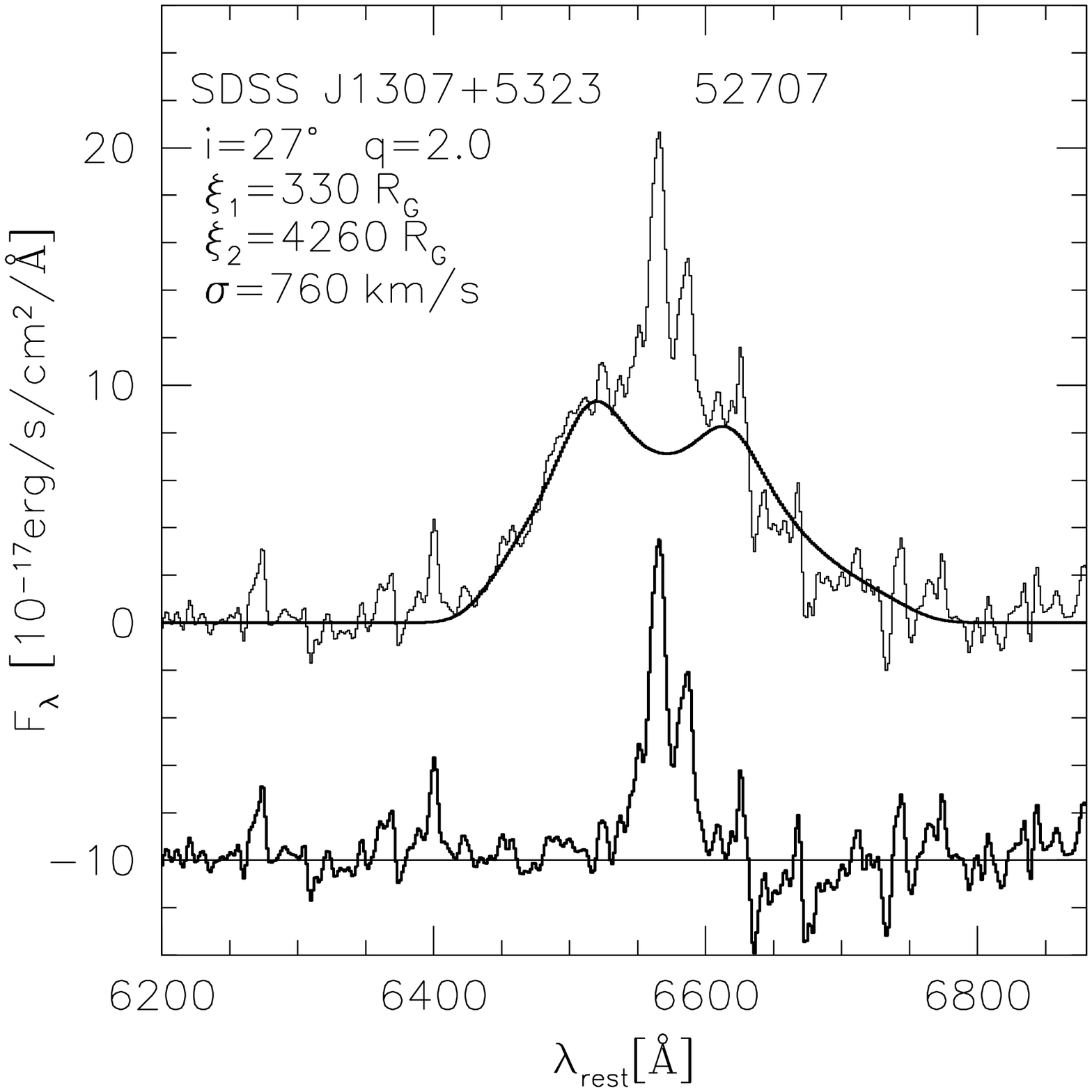}
\caption{Example axisymmetric disk-model fits (solid line overlying the 
spectrum). The residuals, including the narrow H$\alpha$ and
[\ion{N}{2}] lines, are given below, displaced on the flux scale for 
clarity. The SDSS name, the MJD of the observation, and the disk-model
parameters (see \S\ref{fits} for details) are also given in the
top-left corner of each panel.
\label{circFits}} 
\end{figure*}

We attempted to fit all the X-ray observed SDSS AGN with double-peaked
lines with an axisymmetric disk-emission model \citep{CH89}. Six of
the 29 AGN (\about 17\%) allow acceptable axisymmetric disk fits
(i.e. we reject fits with correlated residuals), two of which are
shown in Figure~\ref{circFits}. The emission-region parameters for
these six objects are given in Table~\ref{tab2}. Two double-peaked
emitters, SDSSJ\,0918+5139 and SDSSJ\,1134+4912, with $\xi_1=100\,R_G$
in Table~\ref{tab2}, have poorly constrained inner radii; the only
evidence for a small inner radius in these AGN comes from an extended
blueshifted tail of the relatively narrow H$\alpha$ lines, whose
existence is strongly dependent on the continuum subtraction. Overall,
the parameters listed in Table~\ref{tab2} cover the same ranges as
those found by S03 and Strateva et al. (2006, in preparation) for the
subsample of S03 AGN which allow axisymmetric disk fits. The fraction
of double-peaked emitters with H$\alpha$ line profiles consistent with
emission from an axisymmetric disk is also similar to that found for
the S03 sample Strateva et al. (2006, in preparation).

We conclude that the objects in our main sample are representative of
the S03 sample in their luminosity, redshift, host-starlight
contributions, and broad line-profile parameter distributions.

\subsection{Auxiliary Double-Peaked Sample}

In addition to the X-ray observed double-peaked sample presented
above, we include for comparison purposes 10 RL AGN --- two RL
quasars, four BLRG, and four LINER galaxies --- previously studied in
the X-ray band (see Table~\ref{aux}). Six of these 10 AGN are part of
a recently concluded program to obtain \emph{Hubble Space Telescope}
(HST) UV spectra of nearby RL double-peaked emitters with high-quality
X-ray, optical, and radio observations and a large range of X-ray
luminosities (Eracleous et al. 2006, in preparation). The remaining
four are Arp\,102B, the prototype double-peaked emitter studied in the
X-ray band by \citet{Arp102BXray}, and three low-luminosity LINER
galaxies with archival X-ray observations --- M\,81, NGC\,4203, and
NGC\,4579 (see Table~\ref{aux} and associated references). An
additional BLRG with double-peaked lines from the Eracleous et al.
sample, 3C\,332, was independently found in the SDSS-\emph{ROSAT}
sample, and is considered a main-sample object in this paper.

Eight of these 10 double-peaked emitters have X-ray spectral fits,
including estimates of X-ray power-law photon indices and absorbing
column densities, as well as UV-to-X-ray slopes. When multiple X-ray
observations of an object were available, we used the best
observations available (longest exposures, coverage up to 10\,keV,
lack of pile-up).  Thus for NGC\,4203 we used the photon index from
the spectral fit to the $\sim85$\,ks \emph{ASCA} observation reported
in \citet{ngc4203xray1} ($\Gamma=1.85\pm0.10$), and the X-ray fluxes
and flux ratios obtained from the $\sim1.8$\,ks \emph{Chandra}
observation ($\Gamma=1.9\pm0.3$ in the \hbox{2--10\,keV} region,
assuming no intrinsic absorption). As noted by \citet{ngc4203xray1},
the \emph{ASCA} observation blends the true nuclear flux of NGC\,4203
with that of another object with the same spectral shape. Our more
accurate \emph{Chandra} flux measurement in the \hbox{2--10\,keV} band
agrees to within 20\% with the flux estimate of \citet{ngc4203xray1}
for the nuclear source in NGC\,4203, suggesting that the object was in
a similar flux state. For NGC\,4579 we used a 2001 \emph{XMM-Newton}
observation \citep[see also][]{n4579xmm} and confirmed that the
results did not differ substantially from the 1995 \emph{ASCA}
observation which was obtained only 6 months after the UV data
\citet{n4579asca1}. \citet{n4579asca1} fit the full \emph{ASCA} range
with an absorbed power-law plus a Raymond-Smith model, with a photon
index of $\Gamma=1.72\pm0.05$, absorption consistent with the Galactic
column, and a 2--10\,keV flux of
$4.3\times10^{-12}$\,erg\,cm$^{-2}$\,s$^{-1}$.  Using the 2001
\emph{XMM-Newton} data we obtain a power-law fit with
$\Gamma=1.82\pm0.03$,
$F\si{2--10\,keV}=3.8\pm0.1\times10^{-12}$\,erg\,cm$^{-2}$\,s$^{-1}$,
and no intrinsic absorption. NGC\,4579 was also observed by
\emph{ASCA} in 1998 \citep{n4579asca2} and \emph{Chandra} in 2000
\citep{n4579Chandra} in a higher flux state:
$F\si{2--10\,keV}=5.3\times10^{-12}$\,erg\,cm$^{-2}$\,s$^{-1}$ (1998)
and $F\si{2--10\,keV}=5.2\times10^{-12}$\,erg\,cm$^{-2}$\,s$^{-1}$
(2000), as well as slightly different \hbox{2--10\,keV} photon index
$\Gamma=1.81\pm0.6$ (1998) and $\Gamma=1.88\pm0.03$ (2000).  For M\,81
we used a 2001 \emph{XMM-Newton} pn observation (ObsID 0111800101,
54.3\,ks) and checked that the earliest 1993 \emph{ASCA} observation
\citep{m81ASCA}, which was closer in time to the UV observation, gave
similar results -- $\Gamma=1.81\pm0.02$ and
$F\si{2--10\,keV}=1.4\times10^{-11}$\,erg\,cm$^{-2}$\,s$^{-1}$
(\emph{ASCA}) vs. $\Gamma=1.86\pm0.02$ and
$F\si{2--10\,keV}=1.2\times10^{-11}$\,erg\,cm$^{-2}$\,s$^{-1}$
(\emph{XMM-Newton}; see Table~\ref{aux}).

\begin{deluxetable*}{lccccccclcc} 
\tablewidth{0pt} 
\tablenum{4}
\tabletypesize{\tiny}
\tablecaption{Auxiliary Sample} 
\tablehead{\colhead{Object} & \colhead{$z$} & \colhead{$\Gamma$} & \colhead{$L$(0.1--2.4\,keV)} & \colhead{$L$(2--10\,keV)} & \colhead{$f_{\textrm{2500\,\AA}}$} & \colhead{$\alpha_{\textrm{ox}}$} & \colhead{$l_{\textrm{2500\,\AA}}$} &\colhead{Observatory} &\colhead{Reference} & \colhead{Class}\\
\colhead{(1)} & \colhead{(2)} & \colhead{(3)} & \colhead{(4)} & \colhead{(5)} & \colhead{(6)} & \colhead{(7)} & \colhead{(8)} & \colhead{(9)} & \colhead{(10)} & \colhead{(11)}} 
\startdata
NGC~1097             & 0.0043   & 1.64$\pm$0.03 &  0.041  &  0.00668 & 0.150 & $-$1.22 & 25.763 & \emph{Chandra} & 1 & LINER \\
Pictor~A                 & 0.0350   & 1.80$\pm$0.03 &   96       &  36.0       & 0.340 & $-$0.80 & 27.960 & \emph{ASCA}    & 1 & BLRG \\ 
B2~0742+31         & 0.4610   & 1.80 fixed          & 3200     & ...             & 0.640 & $-$1.30 & 30.688 & \emph{ROSAT}   & 1 & RL quasar\\
PKS~0921$-$213& 0.0531   & 1.74$\pm$0.03 &   54        &  42.5       & 0.370 & $-$1.04 & 28.370 & \emph{XMM}     & 1 & BLRG \\ 
M~81  	                & 0.00086 & 1.89$\pm$0.01 &  0.039   & 0.0175   & 0.310 & $-$0.91 & 24.686 & \emph{XMM}     & 2,3 & LINER \\
NGC~4203            & 0.00362 & 1.85$\pm$0.10 &  0.0190 & 0.0330   & 0.036 & $-$1.02 & 24.993 & \emph{Chandra},\emph{ASCA} & 3,4,5 & LINER \\
NGC~4579           & 0.00507 & 1.82$\pm$0.03 &   0.26     & 0.20     & 0.059 & $-$0.82 & 25.502 & \emph{XMM}     & 3,6 & LINER \\
Arp~102B             & 0.0244   & 1.58$\pm$0.03 &  2.8        &  15.7       & 0.100 & $-$1.02 & 27.108 & \emph{ASCA}    & 1 & BLRG \\ 
PKS~1739+184  & 0.1859    & 1.80$\pm$0.03 &  550       &    ...         & 0.790 & $-$1.25 & 29.867 & \emph{ROSAT}   & 1 & RL quasar\\ 
3C~390.3             & 0.0555    & 1.75$\pm$0.03 &  150       &  139.      & 0.330 & $-$0.87 & 28.361 & \emph{ASCA}    & 1 & BLRG \\ 
\enddata
\tablecomments{(1) Object name; (2) Redshift; (3) \hbox{2--10\,keV}
  power-law slope; (4) \hbox{0.1--2.4\,keV} luminosity in units of
  $10^{42}$\,erg\,s$^{-1}$; (5) \hbox{2--10\,keV} luminosity in units
  of $10^{42}$\,erg\,s$^{-1}$; (6) the 2500\,\AA\ monochromatic flux
  in mJy; (7) the UV-to-X-ray index, $\alpha_{\textrm{ox}}$; (8) the
  logarithm of the 2500\,\AA\ monochromatic luminosity in units of
  erg\,s$^{-1}$\,Hz$^{-1}$; (10) References: 1) Eracleous et al.
  sample. 2) \citet{m81UV} UV spectrum of M~81 3) this work; we used
  the ObsID 397 \emph{Chandra} observation to obtain the fluxes and
  flux ratios (columns 4, 5, and 7) of NGC~4203 \citep[see
  also][]{Ho2001}, the ObsID 0111800101 \emph{XMM-Newton} observation
  for M~81 \citep[see also][]{M81xrays}, and the ObsID 0112840101
  \emph{XMM-Newton} observation for NGC\,4579 \citep[see
  also][]{n4579asca1,n4579asca2,n4579Chandra} 4) \citet{ngc4203UV} UV
  spectrum of NGC~4203 5) \citet{ngc4203xray1} \hbox{2--10\,keV}
  spectral index of NGC~4203.  6) NGC\,4579 UV spectrum from
  \citet{n4579uv}; (11) Optical/radio classification.}
\label{aux}
\end{deluxetable*}

The UV analyses for the auxiliary objects are presented in Eracleous
et al. (2006, in preparation, including the computation of
$\alpha\si{ox}$), \citet{m81UV}, \citet{ngc4203UV}, and
\citet{n4579uv}. The 3C\,332 monochromatic flux, measured at
3700\,\AA, was \about40\% brighter during the SDSS observation than
the earlier HST observation. Since Eracleous et al. have a direct
2500\,\AA\ measurement, we use it instead of the value inferred by
extrapolating the SDSS spectrum. We note that the auxiliary sample of
double-peaked emitters is not representative of the SDSS sample of
double-peaked emitters -- they are all RL, at low redshift, and
predominantly low optical/UV luminosity sources.

\section{X-RAY PROPERTIES OF THE DOUBLE-PEAKED SAMPLE}
\label{sec3}

The presence of X-ray limits (three main-sample double-peaked emitters
and 40 Steffen et al. 2006 RQ AGN) in optically selected samples
requires the use of proper statistical tools when making sample
comparisons. In this section we use the Astronomy SURVival Analysis
package, ASURV \citep{LIF92,IFN86}, to calculate sample means using
the Kaplan-Meier (K-M) estimator, as well as a set of nonparametric
sample-comparison tests for censored data --- the logrank test, Gehan
generalized Wilcoxon test, and Peto \& Prentice generalized Wilcoxon
test (e.g., \S~III of Feigelson \& Nelson 1985 and references
therein). All three tests give better results if the sample sizes are
similar and the censoring distributions are equal. The Peto \&
Prentice test is less vulnerable to different censoring distributions
than the logrank or Gehan tests \citep[see the discussion in \S~5b
of][]{FN85,Latta}. \citet{Fleming} develop a modified Smirnov test for
censored datasets, as the logrank and Gehan tests are insensitive in
the case of ``crossing-hazards alternatives'' (i.e., if the K-M
estimator distribution functions of the different samples are not
close to parallel but cross over instead); this is relevant when
comparing the RL double-peaked emitters with the \citet{Steffen05}
sample discussed below.

\subsection{X-ray Properties of the Double-Peaked Emitters Observed by \emph{ROSAT}}
\label{hrGamma}

To estimate the UV-to-X-ray spectral index,
$\alpha\si{ox}=-0.3838\log[F_{\nu}(\textrm{2500\,\AA})/F_{\nu}(\textrm{2\,keV})]$,
we need a measure of the monochromatic flux at rest-frame 2\,keV.  The
majority (22) of the objects in our sample have been observed only in
the soft X-ray band (\hbox{0.1--2.4\,keV}), and their spectra have
insufficient counts to study their spectral shapes in detail. We first
obtain \hbox{0.5--2\,keV} counts and fluxes, and 2\,keV monochromatic
fluxes (assuming a power-law model with $\Gamma=2$, modified by
Galactic absorption for all sources) following the procedure described
in \citet{aox05} and used by \citet{Steffen05}; we list the results in
Table~\ref{tab3}. The use of a uniform power-law model to determine
the rest-frame 2\,keV monochromatic luminosity and $\alpha\si{ox}$ is
necessary for the purpose of comparison with the \citet{aox05} and
\citet{Steffen05} sources. It is also preferable because of the large
uncertainties of the hardness-ratio-estimated photon indices (which
are also inadequately constrained for 9 of the 22 ROSAT double-peaked
emitters and potentially affected by intrinsic absorption as discussed
below).

\begin{deluxetable*}{cccccccccccccr} 
\tablewidth{0pt} 
\tablenum{5}
\tablecaption{X-ray Properties of the \emph{ROSAT} Subsample} 
\tablehead{\colhead{Object} &\colhead{$z$} &\colhead{Count Rate} &\colhead{Counts} &\colhead{$N_H$} & \colhead{$F$(0.5--2\,keV)} &\colhead{$l_{\textrm{2keV}}$} &\colhead{$l_{\textrm{2500\,\AA}}$}  &\colhead{$\alpha_{\textrm{ox}}$} &\colhead{$\Gamma$} &\colhead{$R$}\\ 
\colhead{(1)} &\colhead{(2)} &\colhead{(3)} &\colhead{(4)} &\colhead{(5)} &\colhead{(6)} &\colhead{(7)} &\colhead{(8)} &\colhead{(9)} &\colhead{(10)} &\colhead{(11)}}
\startdata
022417.17$-$092549.3 &  0.3120 & 0.02240 &   232 &  2.91 &  2.78 & 26.10 & 29.45 & $-$1.29 &  $2.6^{+0.1}_{-0.1 }$ & $<$0.7\\
081329.29$+$483427.9 &  0.2740 & 0.03080 &   133 &  4.58 &  4.03 & 26.13 & 29.29 & $-$1.21 &  $2.0^{+0.4}_{-0.4 }$ & $<$0.6\\
092108.63$+$453857.4 &  0.1740 & 0.09070 &   379 &  1.51 & 10.79 & 26.11 & 28.45 & $-$0.90 &  $1.0^{+0.1}_{-0.1 }$ & 4.0\\
092813.25$+$052622.5 &  0.1870 & 0.01460 &    38 &  3.76 &  1.86 & 25.42 & 28.57 & $-$1.21 &  $<$$2.8^{+0.3}_{-0.7 }$ & $<$0.6\\
094215.13$+$090015.8 &  0.2130 & 0.01120 &    11 &  3.10 &  1.40 & 25.42 & 29.36 & $-$1.51 &  $<$$2.3^{+0.3}_{-0.4 }$ & $<$0.4\\
094745.15$+$072520.6 &  0.0860 & 0.01420 &   152 &  3.01 &  1.77 & 24.66 & 28.39 & $-$1.43 &  $1.6^{+0.2}_{-0.2 }$ & 3.2\\
095427.61$+$485638.1 &  0.2480 & 0.01200 &    26 &  0.98 &  1.40 & 25.57 & 28.97 & $-$1.30 &  $1.6^{+0.5}_{-0.5 }$ & $<$0.9\\
095802.84$+$490311.1 &  0.2420 & 0.01580 &    20 &  0.94 &  1.33 & 25.67 & 29.04 & $-$1.29 &  $2.5^{+0.3}_{-0.7 }$ & $<$0.7\\
104132.78$-$005057.5  &  0.3030 & 0.03340 &   142 &  4.44 &  4.34 & 26.26 & 29.29 & $-$1.16 &  $2.4^{+0.4}_{-0.4 }$ & $<$0.6\\
110109.58$+$512207.1 &  0.2520 & 0.01010 &    12 &  1.09 &  1.19 & 25.52 & 29.17 & $-$1.40 &  $<$$1.5^{+0.3}_{-0.4 }$ & $<$0.9\\
114454.86$+$560238.2 &  0.2310 & 0.01610 &    47 &  1.10 &  1.89 & 25.63 & 28.79 & $-$1.21 &  $2.5^{+0.2}_{-0.3 }$ & $<$0.7\\
114719.22$+$003351.2 &  0.2620 & 0.00420 & $<$16 &  2.28 &  $<$0.51 & $<$25.19 & 29.16 & $<$$-$1.52  & ... & $<$1.1\\
115038.86$+$020854.2 &  0.1090 & 0.00870 &    44 &  2.17 &  1.06 & 24.67 & 27.59 & $-$1.12 &  $<$$1.6^{+0.3}_{-0.3 }$ & $<$0.7\\
115741.94$-$032106.1  &  0.2200 & 0.00440 & $<$11 &  2.32 &  $<$0.54 & $<$25.03 & 29.06 & $<$$-$1.54  & ... & $<$0.6\\
120848.82$+$101342.6 &  0.1160 & 0.05370 &    86 &  1.67 &  6.43 & 25.50 & 27.94 & $-$0.94 &  $2.3^{+0.1}_{-0.1 }$ & $<$0.4\\
130723.13$+$532318.9 &  0.3230 & 0.02700 &   241 &  1.46 &  3.21 & 26.20 & 29.40 & $-$1.23 &  $2.1^{+0.1}_{-0.1 }$ & $<$1.0\\
132355.69$+$652233.0 &  0.2260 & 0.00330 &    16 &  2.01 &  0.39 & 24.93 & 28.65 & $-$1.43 &  $<$$2.4^{+0.3}_{-0.4 }$ & $<$1.1\\
144302.76$+$520137.2 &  0.1410 & 0.12040 &   787 &  1.63 & 14.38 & 26.04 & 28.44 & $-$0.92 &  $1.7^{+0.1}_{-0.1 }$ & 3.6\\
161742.53$+$322234.3 &  0.1510 & 0.00870 &    84 &  2.02 &  1.05 & 24.96 & 28.96 & $-$1.53 &  $<$$1.2^{+0.3}_{-0.3 }$ & 3.3\\
164031.86$+$373437.2 &  0.2800 & 0.02010 &    60 &  1.32 &  2.38 & 25.92 & 28.74 & $-$1.08 &  $2.3^{+0.2}_{-0.2 }$ & $<$1.1\\
170102.29$+$340400.6 &  0.0950 & 0.03100 &   140 &  2.06 &  3.75 & 25.08 & 28.24 & $-$1.21 &  $1.9^{+0.1}_{-0.1 }$ & $<$0.3\\
213338.42$+$101923.6 &  0.1260 & 0.00420 & $<$16 &  4.83 &  $<$0.55 & $<$24.51 & 28.36 & $<$$-$1.48  & ... & $<$0.9\\
\enddata
\tablecomments{(1) the SDSS name given in the J2000 epoch RA and Dec
  form, HHMMSS.ss$\pm$DDMMSS.s; (2) redshift; (3) PSPC
  \hbox{0.5--2\,keV} count rate, in counts per second; (4) total
  counts in the \hbox{0.5--2\,keV} band; (5) Galactic absorbing
  column, in units of $10^{20}$\,cm$^{-2}$; (6) flux in the
  \hbox{0.5--2\,keV} band (assuming a power-law spectrum with
  $\Gamma=2$), corrected for Galactic absorption, in units of
  $10^{-13}\,$\,erg\,cm$^{-2}$\,s$^{-1}$; (7) the logarithm of the
  2\,keV monochromatic luminosity in units of
  erg\,s$^{-1}$\,Hz$^{-1}$; (8) the logarithm of the 2500\,\AA\
  monochromatic luminosity in units of erg\,s$^{-1}$\,Hz$^{-1}$; (9)
  the UV-to-X-ray spectral index, $\alpha_{\textrm{ox}}$; (10) the
  hardness-ratio estimate of the spectral slope, $\Gamma$; (11) $R$ is
  the radio-loudness indicator, $R=\log(F_{\textrm{1.4\,GHz}}/F_i)$,
  where $F_i$ is the SDSS $i$-band flux and $R \geq 1$ is considered
  radio loud. }
\label{tab3}
\end{deluxetable*}

In addition to the above analysis, we also estimate the spectral shape
of the \emph{ROSAT}-observed main-sample AGN using the standard
hardness ratio, HR1, defined as $\textrm{HR1}=(B-A)/(B+A)$, where $A$
and $B$ represent the total numbers of photons in the
\hbox{0.11--0.41\,keV} and \hbox{0.5--2\,keV} bands, respectively.
Assuming no intrinsic absorption above the Galactic value, and spectra
well represented by a power-law models, we can derive a one-to-one
relationship between the power-law photon index $\Gamma$ and HR1. We
used
PIMMS\footnote{http://heasarc.gsfc.nasa.gov/docs/software/tools/pimms.html}
to create a grid of expected HR1 values assuming an absorbed power-law
model with different input photon indices and absorbing Galactic
column densities spanning the observed range
($10^{20}\,\textrm{cm}^{-2}<N_H<5\times10^{20}\,\textrm{cm}^{-2}$).
We found cubic polynomial relations between $\Gamma$ and HR1 for each
absorbing column and used these to estimate the photon index
associated with each observed HR1, $N_H$ pair. To estimate the
photon-index uncertainties, we propagated uncertainties of the count
rates in the \hbox{0.5--2\,keV} and \hbox{0.11--0.41\,keV}
\emph{ROSAT} bands using the method described in \S1.7.3 of
\citet{lyons}. The results are given in Table~\ref{tab3}. The three
double-peaked emitters, which are undetected in the \hbox{0.5--2\,keV}
band, have no HR1 photon index estimates. Six of the remaining 19
double-peaked emitters detected by \emph{ROSAT} were detected in the
\hbox{0.5--2\,keV} band only; for these we can estimate only an HR1
lower limit and a corresponding upper limit on the power-law slope
$\Gamma$.  Any intrinsic absorption in excess of Galactic will
translate into an underestimate of the true value of $\Gamma$. This
fact is of particular significance for the six double-peaked emitters
without \hbox{0.11--0.41\,keV} detections, since the upper-limit
values of $\Gamma$ reported in Table~\ref{tab3} could be
underestimated, rendering them useless for comparison. In
\S~\ref{Sgamma} we point out that this could indeed be the case for a
fraction of the AGN in the current sample.

\subsection{X-ray Properties of the Double-Peaked Emitters Observed by 
\emph{XMM-Newton} and \emph{Chandra}} 
\label{spGamma}

For the seven main-sample double-peaked emitters with hard-band
coverage we were able to measure the monochromatic flux at 2\,keV
(rest-frame) directly from the spectrum and report the corresponding
2\,keV monochromatic luminosities, together with their UV-to-X-ray
indices in Table~\ref{tab6}. These 2\,keV measurements might be
underestimates of the true values if intrinsic absorption above
$10^{22}\,\textrm{cm}^{-2}$ (assuming $z\approx0.2$) is present.  In
the analysis of individual objects below, we show that this is
probably the case for only two of the \emph{XMM-Newton} or
\emph{Chandra} observed objects and we correct the 2\,keV measurements
accordingly.

\begin{deluxetable*}{ccccccccccccr} 
\tablewidth{0pt} 
\tablenum{6}
\tablecaption{X-ray Properties of the \emph{XMM-Newton}/\emph{Chandra} Subsample} 
\tablehead{\colhead{Object} &\colhead{$z$} &\colhead{$l_{\textrm{2500\,\AA}}$} &\colhead{$l_{\textrm{2\,keV}}$} &\colhead{$\alpha_{\textrm{ox}}$} &\colhead{F(2--10\,keV)}  &\colhead{$\Gamma$} & \colhead{Fit Range} &\colhead{Model} &\colhead{$N_H$} &\colhead{$N_{H,\textrm{intr}}$} &\colhead{Counts} & \colhead{$R$} \\
\colhead{(1)} &\colhead{(2)} &\colhead{(3)} &\colhead{(4)}  &\colhead{(5)} &\colhead{(6)} &\colhead{(7)} &\colhead{(8)} &\colhead{(9)}  &\colhead{(10)} &\colhead{(11)} &\colhead{(12)} &\colhead{(13)}}
\startdata
0043$+$0051  & 0.3081 & 29.76 & 26.53 & $-$1.24	& 9.3   & 1.70$\pm$0.06  & 2--10	& PL   & 2.31 & ...   & 1704   & 0.8 \\ 
0057$+$1446  & 0.1722 & 30.03 & 26.55 & $-$1.33	& 38.0 & 2.2$\pm$0.4     & 1--10	& PL    & 4.37 & ...   & 403    & $<$0.0\\
0918$+$5139  & 0.1854 & 29.18 & 25.33 & $-$1.48	& 2.9   & 1.4$\pm$0.3     & 0.4--11 & PAPL & 1.45 & $500\pm100$  & 243   & $<$0.3\\
...                        & 0.1854 & 29.18 & 25.05 & $-$1.58	& 2.0   & 1.3$\pm$0.4     & 0.3--11	& APL & 1.45 & $130\pm60$     & 145   & $<$0.3\\
0938$+$0057  & 0.1704 & 28.95 & 25.95 & $-$1.15	& 18.2 & 1.5 fixed           & 0.3--11 	& APL & 4.22 & $200\pm40$  & 132    & $<$$-$0.3\\
1111$+$4820  & 0.2809 & 29.49 & 25.73 & $-$1.44	& 2.6   & 1.7$\pm$0.2     & 2--10 	& PL    & 1.24 & ...   & 295    & 1.7 \\	
1134$+$4912  & 0.1765 & 28.57 & 24.46 & $-$1.58	& $>$0.25 & 1.9 fixed     & 0.2--10 	& PL    & 1.55 & ...   & 24     & 0.8 \\	
2304$-$0841   & 0.0469 & 28.59 & 26.19 & $-$0.92	& 302. & 1.71$\pm$0.02 & 2--10 	& PLG & 3.60 & ...   & 27200  & 1.0 \\	
\enddata
\tablecomments{(1) the SDSS name given in the J2000 epoch RA and Dec
  form, HHMMSS.ss$\pm$DDMMSS.s; (2) redshift; (3) the logarithm of the
  rest-frame 2500\,\AA\ monochromatic luminosity in units of
  erg\,s$^{-1}$\,Hz$^{-1}$; (4) the logarithm of the rest-frame 2\,keV
  monochromatic luminosity in units of erg\,s$^{-1}$\,Hz$^{-1}$; (5)
  the UV-to-X-ray spectral index, $\alpha_{\textrm{ox}}$; (6) the
  unabsorbed flux in the observed \hbox{2--10\,keV} band in units of
  $10^{-13}$\,erg\,cm$^{-2}$\,s$^{-1}$; (7) the spectral photon index,
  $\Gamma$; (8) the range of the spectral fit, in keV; (9)
  spectral-fit models: PL corresponds to a simple power-law model
  including Galactic absorption (i.e. ``wabs pow'' in XSPEC); APL
  refers to a intrinsically absorbed power-law model (``wabs zwabs
  pow)'' in XSPEC), with the column density of the intrinsic
  absorption given in column (11); PAPL refers to a partially absorbed
  power-law model (``wabs zpcfabs pow)'' in XSPEC); PLG refers to a
  power-law model with galactic absorption and a Gaussian Fe-K$\alpha$
  line (``wabs pow zgaus'' in XSPEC); (10) The Galactic absorption at
  the AGN position in units of $10^{20}$\,cm$^{-2}$; (11) any
  intrinsic absorption in units of $10^{20}$\,cm$^{-2}$; (12) the
  total counts (\emph{Chandra} ACIS-S or \emph{XMM-Newton} pn, except
  for SDSSJ\,1134$+$4912, where the \emph{XMM-Newton} MOS counts are
  given) in the energy range given in column (8); (13) the radio
  loudness, $R=\log(F_{1.4GHz}/F_i)$.}
\label{tab6}
\end{deluxetable*}

Five of the double-peaked emitters have sufficient counts to obtain
absorbed power-law fits using
XSPEC.\footnote{http://heasarc.gsfc.nasa.gov/docs/xanadu/xspec/} For
the remaining two objects we attempt to constrain the intrinsic
absorption assuming a standard power-law photon index. In all cases we
include the Galactic absorption in the model fits. The results of the
analysis for all seven double-peaked emitters with hard-band X-ray
detections are presented below.

\subsubsection{SDSSJ\,0043$+$0051 (UM\,269)}
\label{spec2}

SDSSJ\,0043$+$0051 ($z=0.3081$) has been observed by
\emph{XMM-Newton}, \emph{ASCA}, and \emph{ROSAT} (see
Table~\ref{tab1}).  The \emph{XMM-Newton} spectral fits were published
by \citet{um296spec}; we confirm their results: the observed
\hbox{2--10\,keV} spectrum can be represented by a power law with
photon index $\Gamma=1.70\pm0.06$ and Galactic absorption
($\chi^2/DoF=0.9$ for 79\,$DoF$, EPIC pn fit). The \hbox{0.5--10\,keV}
spectrum and fit are shown in Figure~\ref{Xspectra}. When this model
is extended to lower energies, a soft-excess is detected. There is no
evidence of Fe~K$\alpha$ emission; for a line centered at 6.4\,keV
with a width of 10\,eV, we obtain an upper limit of 180\,eV for the
equivalent width (EW) at 90\% confidence (Page et al. 2004 estimate
EW$<80$\,keV at 90\% confidence). For comparison, an \emph{ASCA}
power-law fit for a 31\,ks observation obtained five years earlier (as
reported in the Tartarus database at HEASARC\footnote{High Energy
  Astrophysics Science Archive Research Center,
  http://heasarc.gsfc.nasa.gov/}) has $\Gamma=1.4\pm0.1$, a 60\%
higher flux in the \hbox{2--10\,keV} band,
$F\si{2--10\,keV}=1.6\times10^{-12}$\,erg\,cm$^{-2}$\,s$^{-1}$,
and no evidence for Fe~K$\alpha$ emission.

\subsubsection{SDSSJ\,0057$+$1446}
\label{spec3}

The observed \emph{Chandra} ACIS-S count rate per frame for
SDSSJ\,0057+1446 ($z=0.1722$) is $\sim0.57$ counts s$^{-1}$ (with
frame-time of 0.441\,s) and is high enough for photon pile-up to be
important even for an off-axis angle of $3.3'$.  Pile-up is the
incidence of two or more X-ray photons in one (or more) neighboring
CCD pixels within one frame time. The CCD electronics may falsely
regard these events as a single event with an amplitude given by the
sum of the electron charge, resulting in a decrease of the apparent
count rate of the source, an artificial hardening of its spectrum, and
the apparent distortion of the point-spread function (PSF) of
point-like objects.  It can also alter the photon event grades and
lead to a loss of events when standard grade filtering is applied to
the data.

To account for the effects of pile-up, we initially analyzed spectra
of SDSSJ\,0057+1446 from annuli of various inner radii --- 0$''$,
4$''$, 1.0$''$, 1.5$''$ , and 2$''$ --- and an outer radius of
3.4$''$, excluding events from the core of the PSF in annuli \#2
through \#5, which are most affected by pile-up.  The effective-area
files used when modeling the annular spectra were corrected for the
energy dependence of the \emph{Chandra} PSF. We fit the annular
spectra with simple power-law models modified by Galactic absorption.
The resulting \hbox{1--10\,keV} spectral slopes are
$\Gamma=1.86\pm0.08$ (annulus \#1), $\Gamma=1.9\pm0.1$ (\#2),
$\Gamma=2.0\pm0.2$ (\#3), and $\Gamma=2.2\pm0.4$ (\#4), with no need
for absorption above Galactic. The last annulus (\#5) has very few
photons resulting in a poorly constrained spectral index,
$\Gamma=2\pm1$. Based on the spectral fit in annulus \#4
(1.5--3.4$''$) which is shown in Figure~\ref{Xspectra}, we estimate
the pile-up corrected $\Gamma$ = $2.2\pm0.4$.  To estimate the
fraction of events lost due to pile-up we used the forward
spectral-fitting tool LYNX that simulates pile-up (Chartas et al.
2000). LYNX simulates the propagation of individual photons through
the \emph{Chandra} mirrors and ACIS, and it takes into account the
possible overlap of events within one frame time.  We find that the
pile-up fraction is $\sim$ 20\% in the \hbox{0.5--2\,keV} band (the
pile-up corrected flux is
$3.1\times10^{-12}$\,erg\,cm$^{-2}$\,s$^{-1}$) and is negligible in
the \hbox{2--10\,keV} band
($F\si{2--10\,keV}=3.8\times10^{-12}$\,erg\,cm$^{-2}$\,s$^{-1}$).

\subsubsection{SDSSJ\,0918$+$5139}
\label{spec4}

SDSSJ\,0918$+$5139 (\hbox{$z=0.1854$}) was serendipitously observed by
both \emph{Chandra} and \emph{XMM-Newton} (see Table~\ref{tab6}) as a
result of its proximity (\about 6 arcmin) to a nearby cluster,
Abell\,773 (\hbox{$z=0.21$}). The \emph{XMM-Newton} pn spectrum has 243
counts in the \hbox{2--11\,keV} band. A power-law fit with no
intrinsic absorption above the Galactic value provides a marginally
acceptable fit in the \hbox{2--11\,keV} band -- $\chi^2/DoF\approx1.5$
for 15\,$DoF$ with \hbox{$\Gamma=0.6\pm0.2$}.  The \hbox{2--11\,keV}
fit can be improved by adding intrinsic absorption --
$\chi^2/DoF\approx1.1$ for 14\,$DoF$, with \hbox{$\Gamma=1.7\pm0.5$}
and $N\si{H,intr}=(8\pm4)\times10^{22}$\,cm$^{-2}$, but neither of these
fits is acceptable over the full \hbox{0.4--11\,keV} \emph{XMM-Newton}
band (the best fit has $\chi^2/DoF\approx1.7$, 21\,$DoF$).  A model
with a partial absorber in addition to the Galactic one --
\hbox{$\Gamma=1.4\pm0.3$},
$N\si{H,intr}=(5\pm1)\times10^{22}$\,cm$^{-2}$, and a covering fraction
of $f\si{c}=0.92\pm0.05$ -- provides an acceptable fit in the
\hbox{0.4--11\,keV} band: $\chi^2/DoF\approx1.0$ for 20\,$DoF$.  The
\emph{Chandra} spectrum has only 102 counts in the \hbox{2--10\,keV}
region and can be represented by a \hbox{$\Gamma=1.5\pm0.4$} power law
fit with no intrinsic absorption above the Galactic value
($\chi^2/DoF\approx1.1$, 17\,$DoF$). The full \emph{Chandra} band
(\hbox{0.3--11\,keV}, 145\,counts) requires an intrinsic absorber
($\chi^2/DoF\approx1.5$ vs. $\chi^2/DoF\approx1.1$, 24\,$DoF$)
$N\si{H,intr}=(1.3\pm0.6)\times10^{22}$\,cm$^{-2}$ and has a slightly
flatter spectral slope, \hbox{$\Gamma=1.3\pm0.4$}. The data suggests
that the intrinsic absorber has changed in the seven months separating the
\emph{XMM-Newton} and \emph{Chandra} observations. In the following
sections (\S~\ref{Saox} and \S~\ref{Sgamma}) we use the
\hbox{0.4--11\,keV} \emph{XMM-Newton} results, which were obtained 4.6
months after the optical spectrum, have better signal-to-noise, and
more accurate fit parameters.

The absorption corrected \hbox{2--10\,keV} fluxes are consistent
between the two observations within the (90\% confidence) errorbars:
$F\si{2--10\,keV}=2.0_{-0.4}^{+0.7}\times10^{-13}$\,erg\,cm$^{-2}$\,s$^{-1}$
(\emph{Chandra}), and
$F\si{2--10\,keV}=2.9_{-1.2}^{+0.8}\times10^{-13}$\,erg\,cm$^{-2}$\,s$^{-1}$
(\emph{XMM-Newton} pn).  In the soft band, the observed fluxes are
$F\si{0.5--2\,keV}=1.8_{-1.5}^{+0.7}\times10^{-14}$\,erg\,cm$^{-2}$\,s$^{-1}$
(\emph{Chandra}) and
$F\si{0.5--2\,keV}=0.9_{-0.7}^{+0.4}\times10^{-14}$\,erg\,cm$^{-2}$\,s$^{-1}$
(\emph{XMM-Newton} pn); the intrinsic absorption correction could
increase those by a factor of 3 (\emph{Chandra}) to 6 (\emph{XMM-Newton}).

\subsubsection{SDSSJ\,0938$+$0057}
\label{spec5}

SDSSJ\,0938$+$0057 ($z=0.1704$) has a 132 count detection
(\hbox{0.3--10\,keV}) in a 1.3\,ks \emph{Chandra} observation. The
spectrum is hard, implying high intrinsic obscuration and/or a flat
power-law. If we fix the absorbing column to the Galactic value,
$N_H=4.2\times10^{20}$\,cm$^{-2}$, a simple power-law fit in the
\hbox{2--10\,keV} band (87\,counts) returns $\Gamma=0.9\pm0.4$.
Alternatively, assuming a standard power-law slope, $\Gamma=1.9$ for
RQ AGN, we infer an intrinsic absorbing column of
$N\si{H,intr}=(6.4\pm2.3)\times10^{22}$\,cm$^{-2}$. Neither of these fits is
applicable over the full \hbox{0.3--10\,keV} range: the required
photon index drops to $\Gamma=0.1\pm0.1$ if we assume no intrinsic
absorption, which is unrealistic considering the uniformity of the
power-law slopes measured in RQ AGN \citep[$1.9\pm0.5$, see, for
example, Figure 6 of][ and references therein]{Gref}, and the best
$\Gamma=1.9$ model including intrinsic absorption is inadequate
($\chi^2/DoF\approx1.6$, 24\,$DoF$). The full \hbox{0.3--10\,keV} band
spectrum can be represented by an intrinsically absorbed model with
fixed $\Gamma=1.5$ and an intrinsic absorbing column of
$N\si{H,intr}=(2.0\pm0.4)\times10^{22}$\,cm$^{-2}$ ($\chi^2/DoF\approx1.3$,
24\,$DoF$). For a simple intrinsically absorbed power law model,
photon indices steeper than $\Gamma=1.5$ are excluded. If we allow an
intrinsic absorber with partial coverage, the intrinsic absorbing
column increases $N\si{H,intr}=(5\pm1)\times10^{22}$\,cm$^{-2}$, allowing a
$\Gamma=1.8$ fit ($\chi^2/DoF\approx1.1$, 24\,$DoF$). Higher
signal-to-noise data is necessary to distinguish between the flat
photon index, intrinsically absorbed model, and the steep photon
index, partially absorbed model.

We adopt the simpler \hbox{0.3--10\,keV} band $\Gamma=1.5$ model fit
(see Table~\ref{tab6} and Figure~\ref{Xspectra}) and estimate the
unabsorbed fluxes in the hard and soft bands:
$F\si{2--10\,keV}=1.3^{+0.3}_{-0.7}\times10^{-12}$\,erg\,cm$^{-2}$\,s$^{-1}$
and
$F\si{0.5--2\,keV}=1.2^{+0.2}_{-0.4}\times10^{-12}$\,erg\,cm$^{-2}$\,s$^{-1}$.

\subsubsection{SDSSJ\,1111$+$4820}
\label{spec6}

\begin{figure*}
\epsscale{1.1}
\plotone{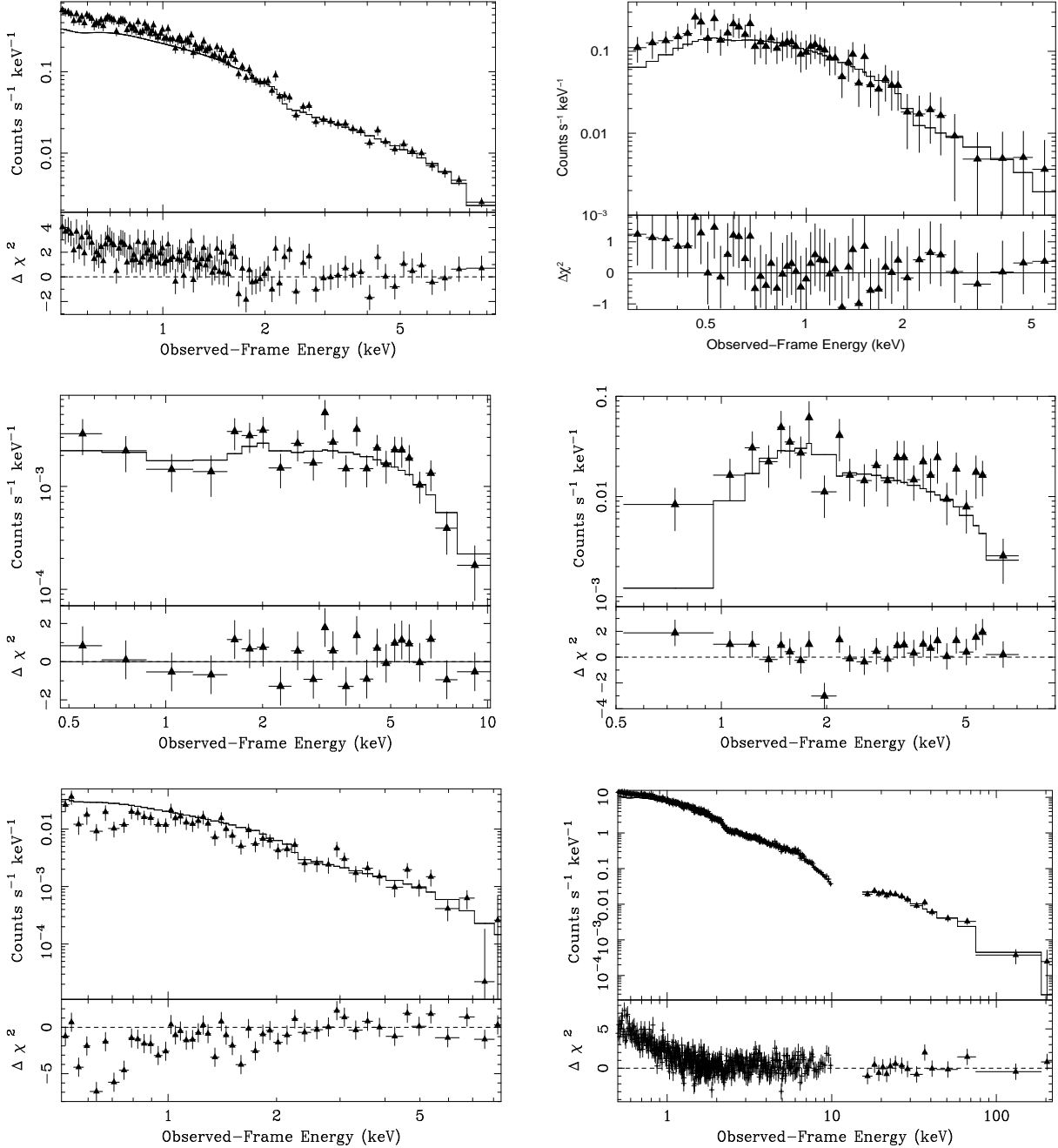}
\caption{X-ray spectra and model fits of five main-sample
double-peaked emission AGN: SDSSJ\,0043+0051 (\emph{XMM-Newton} pn
spectrum, top left), SDSSJ\,0057+1446 (\emph{Chandra} spectrum of
annulus \#4, top right), SDSSJ\,0918+5139 ( \emph{XMM-Newton} pn spectrum,
middle left), SDSSJ\,0938+0057 (\emph{Chandra} spectrum, middle right),
SDSSJ\,1111+4820 (\emph{XMM-Newton} pn spectrum, bottom left), and
SDSSJ\,2304$-$0841 (simultaneous \emph{XMM-Newton} and
\emph{Beppo-SAX} data, bottom right).
\label{Xspectra}}
\end{figure*}

SDSSJ\,1111$+$4820 ($z=0.2809$) falls within the field of view of two
\emph{XMM-Newton} observations, one of which (Obs. ID 0059750401) is
heavily affected by background flaring. We used the pn observation
from June 2002 (Obs. ID 0104861001), which has a much longer effective
exposure time, to extract and fit the \hbox{2--10\,keV} spectrum of
SDSSJ\,1111$+$4820. The results are given in Figure~\ref{Xspectra} and
in Table~\ref{tab6}. The \hbox{2--10\,keV} spectrum is well fit by a
simple power law with Galactic absorption with $\Gamma=1.7\pm0.2$. If
we extend this model to lower energies, the negative residuals below
2\,keV could indicate the presence of intrinsic absorption.  Excess
intrinsic absorption, however, is not required by the data: the
\hbox{0.5--10\,keV} spectrum also admits a power-law fit with a
flatter photon index -- $\Gamma=1.35\pm0.07$ -- and no intrinsic
absorption ($\chi^2/\textrm{DoF}=1.2$ for 48\,$DoF$).

\subsubsection{SDSSJ\,1134$+$4912}
\label{blended}

SDSSJ\,1134$+$4912 ($z=0.1765$) is situated less than 10$''$ from
another, optically fainter, point-like object in the SDSS image
(SDSS\,J113451.64$+$491200.8, with an $r$-band magnitude of
21.8). Both this fainter object and the double-peaked emitter of
interest to us are detected separately in the \emph{XMM-Newton} MOS
images with similar counts; the larger pixel size of the pn (4.1$''$
for pn vs. 1.1$''$ for the MOS detectors) blends the two objects, with
a centroid closer to the fainter MOS object. The \emph{ROSAT} PSPC
(PSF FWHM of \about25$''$ at 1\,keV and an observed off-axis angle of
$\sim9'$) or HRI (with a 50\% power radius of \about9$''$ for an
off-axis angle of $\sim12'$) lack the angular resolution to separate
the two objects clearly and will not be considered here. The measured
0.5--2\,keV flux of the double-peaked emitter is
$F\si{0.5--2\,keV}=(2\pm1)\times10^{-14}$\,erg\,cm$^{-2}$\,s$^{-1}$
for a 6$''$ aperture (MOS2). Using the EPIC MOS on-axis PSF (which
does not change much for objects at an off-axis angle of 7$'$ for
energies $<$2\,keV), we estimate that the total flux of
SDSSJ\,1134$+$4912 is
$F\si{0.5--2\,keV} \approx 3\times10^{-14}$\,erg\,cm$^{-2}$\,s$^{-1}$. On
account of the proximity to the second source, the small aperture used
for the flux estimate, and the small number of total counts, the X-ray
flux and 2\,keV monochromatic luminosity of this object are
uncertain. Using the Cash statistic, we can obtain an acceptable fit
to the MOS spectrum using a power-law model with $\Gamma=1.9$ and an
absorbing column density equal to the Galactic value. The high
resolution of \emph{Chandra} is necessary to measure the X-ray
properties of this object properly.

\subsubsection{SDSSJ\,2304$-$0841 (MCG-2-58-22, Mrk\,926)}
\label{spec1}

SDSSJ\,2304$-$0841 is a nearby ($z=0.0469$) Seyfert galaxy that has
been observed repeatedly with most current and past X-ray
observatories (see Appendix~\ref{sec:app}). Figure~\ref{Xspectra}
shows the spectrum in the 0.5--220\,keV range, originally published by
\citet{926spec}, overlaid with a model consisting of an exponentially
cutoff power-law, reflection from an isotropically illuminated cold
slab \citep{pexrav}, and a Gaussian Fe-K$\alpha$ line. Our 2--220\,keV
XSPEC fits confirm their results: the spectrum above 2\,keV is well
represented by an absorbed and Compton reflected power law with
$\Gamma=1.76\pm0.04$, absorption equal to the Galactic column density,
a 6.4\,keV Fe~K$\alpha$ line with $EW=60_{-40}^{+80}$\,keV and energy
dispersion $\sigma=(120\pm80)$\,eV (corresponding to
FWHM$\approx13000$\,km\,s$^{-1}$, comparable to that of the H$\alpha$
line, FWHM$\approx11400$\,km\,s$^{-1}$), and a soft excess below
2\,keV. If we consider only the \hbox{2--10\,keV} region (see
Table~\ref{tab6}), the power law fit with no intrinsic absorption and
a Gaussian Fe~K$\alpha$ line has $\Gamma=1.71\pm0.02$, consistent with
the value obtained above in the 0.5--220\,keV range. Three previous
spectra observed with \emph{ASCA} obtained in 1993 and 1997
\citep{asca93,mkn926FeKa} show similar spectral shapes, with
$1.71<\Gamma<1.84$ and evidence for some intrinsic absorption
($N\si{H,intr} \approx10^{21}$\,cm$^{-2}$) in the two 1997 spectra.
Judging from the relation between X-ray and optical variability, which
is detailed in Appendix~\ref{sec:app}, the presence of intrinsic
absorption seems unrelated to the shape of the H$\alpha$ line. Both
the 1993 and one of the 1997 spectra show clear Fe~K$\alpha$ lines,
whose parameters depend on the continuum model
\citep{asca93,mkn926FeKa}; the presence of a strong Fe~K$\alpha$ line
also appears unrelated to the single- or double-peaked shape of the
optical Balmer lines (Appendix~\ref{sec:app}).

\begin{figure}
\epsscale{1.0}
\plotone{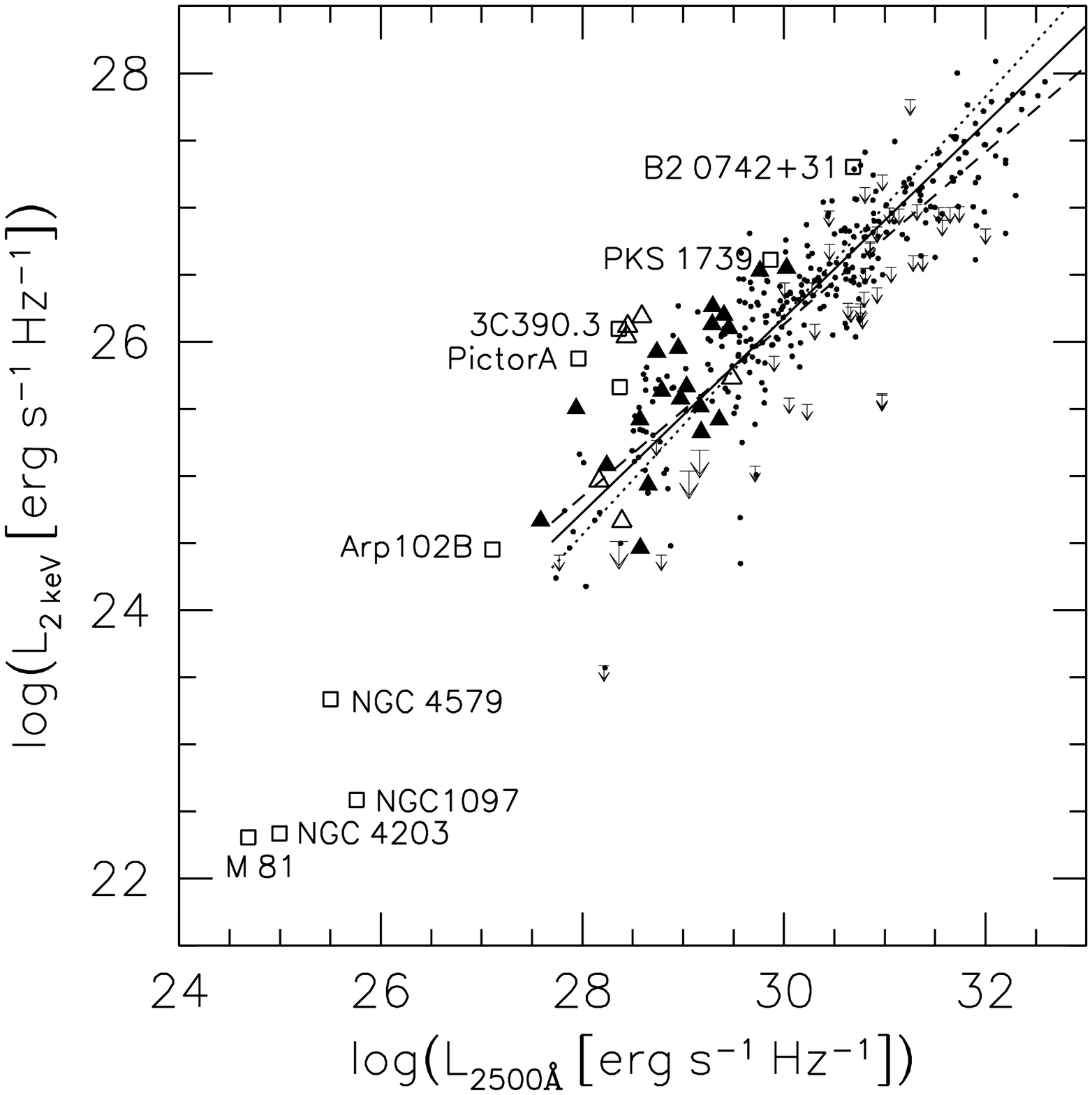}
\caption{Rest-frame UV vs. X-ray monochromatic luminosities of the
main SDSS double-peaked sample (open triangles denote RL AGN, solid
triangles denote RQ AGN, and large arrows denote X-ray limits), the
\citet{Steffen05} RQ AGN sample (dots and small arrows, indicating
X-ray detections and upper limits, respectively), and the auxiliary
double-peaked sample (open squares). The solid line is the best-fit
bisector linear regression from \citet{Steffen05}; the dashed and
dotted lines represent the fits minimizing the residuals in ordinate
and the abscissa, respectively (see Steffen et al. 2006 for more
details).  The difference between the dashed and dotted lines can be
used as an indication of the maximum uncertainly in the UV-X-ray
relation.
\label{lxluv}} 
\end{figure}

\subsection{UV-to-X-ray Slope}
\label{Saox}

The rest-frame UV and soft X-ray emission from AGN are correlated,
with more luminous AGN emitting relatively less X-rays per unit UV
luminosity \citep[e.g.,][ and references
therein]{aox05,Steffen05}. Consequently a proper comparison of the
X-ray emission of double-peaked emitters with those of normal AGN
should take this relation into account.  Figure~\ref{lxluv} shows the
2500\,\AA\ vs. the 2\,keV monochromatic luminosities of the
double-peaked emitters in comparison with those of RQ AGN from the
\citet{Steffen05} sample. The \citet{Steffen05} sample includes 333
optically selected RQ AGN with a high X-ray detection fraction (88\%),
and no evidence of intrinsic absorption in the UV. Over half of the
objects in the sample are SDSS AGN, directly comparable to the
double-peaked emitters studied here. The remaining objects allow us to
extend the UV luminosity range of the comparison sample at each
redshift. The RQ double-peaked emitters appear indistinguishable in
Figure~\ref{lxluv} from normal AGN with similar UV monochromatic
luminosities. The majority of the RL double-peaked emitters show
excess 2\,keV emission, as is characteristic of all RL AGN
\citep[e.g.,][]{XrayRadio}.

\begin{figure}
\epsscale{1.0}
\plotone{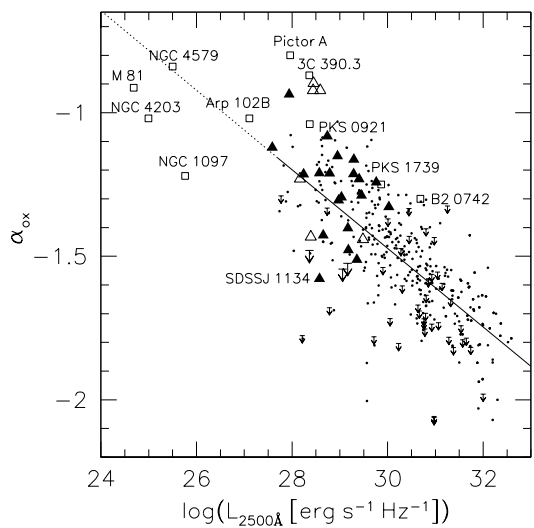}
\caption{Rest-frame UV monochromatic luminosity vs.  $\alpha$\si{ox}.
  The symbols are as in Figure~\ref{lxluv}. The line is the best-fit
  linear regression from \protect \citet{Steffen05} with an
  extrapolation to lower luminosities given by the dotted line.
\label{aox1}}
\end{figure}

Figure~\ref{aox1} shows the UV-to-X-ray slope, $\alpha$\si{ox}, for
the double-peaked emitters in comparison with the sample of
\citet{Steffen05}.  The majority of double-peaked AGN follow the
\citet{Steffen05} $\alpha\si{ox}$-$l\si{2500\,\AA}$ relation,
$\alpha\si{ox}(l\si{2500\,\AA})=-0.137l\si{2500\,\AA}+2.638,$ and do
not differ substantially from normal AGN with comparable luminosity in
the UV. Due to the small range of luminosities probed at each redshift
and the strong luminosity-redshift relation in the sample of
double-peaked emitters, a partial-correlation analysis for the
$\alpha\si{ox}$-$l\si{2500\,\AA}$ relation (i.e., one that takes into
account the $l\si{2500\,\AA}$-$z$ relation) is inconclusive.  The
strength of the $\alpha\si{ox}$-$l\si{2500\,\AA}$ partial correlation
is only 2.3$\sigma$ with partial Kendall's $\tau_{12,3}=-0.24$, which
could indicate either an inability of the partial-correlation analysis
to detect a weaker $\alpha\si{ox}$-$l\si{2500\,\AA}$ anti-correlation
on top of a strong $l\si{2500\,\AA}$-$z$ correlation
($\tau_{23}=0.53$), or a genuine lack of an
$\alpha\si{ox}$-$l\si{2500\,\AA}$ relation. Since we know the
$l\si{2500\,\AA}$-$z$ correlation is very strong for our sample of
double-peaked emitters, we believe the former is more likely.

\begin{figure*}
\epsscale{1.0}
\plottwo{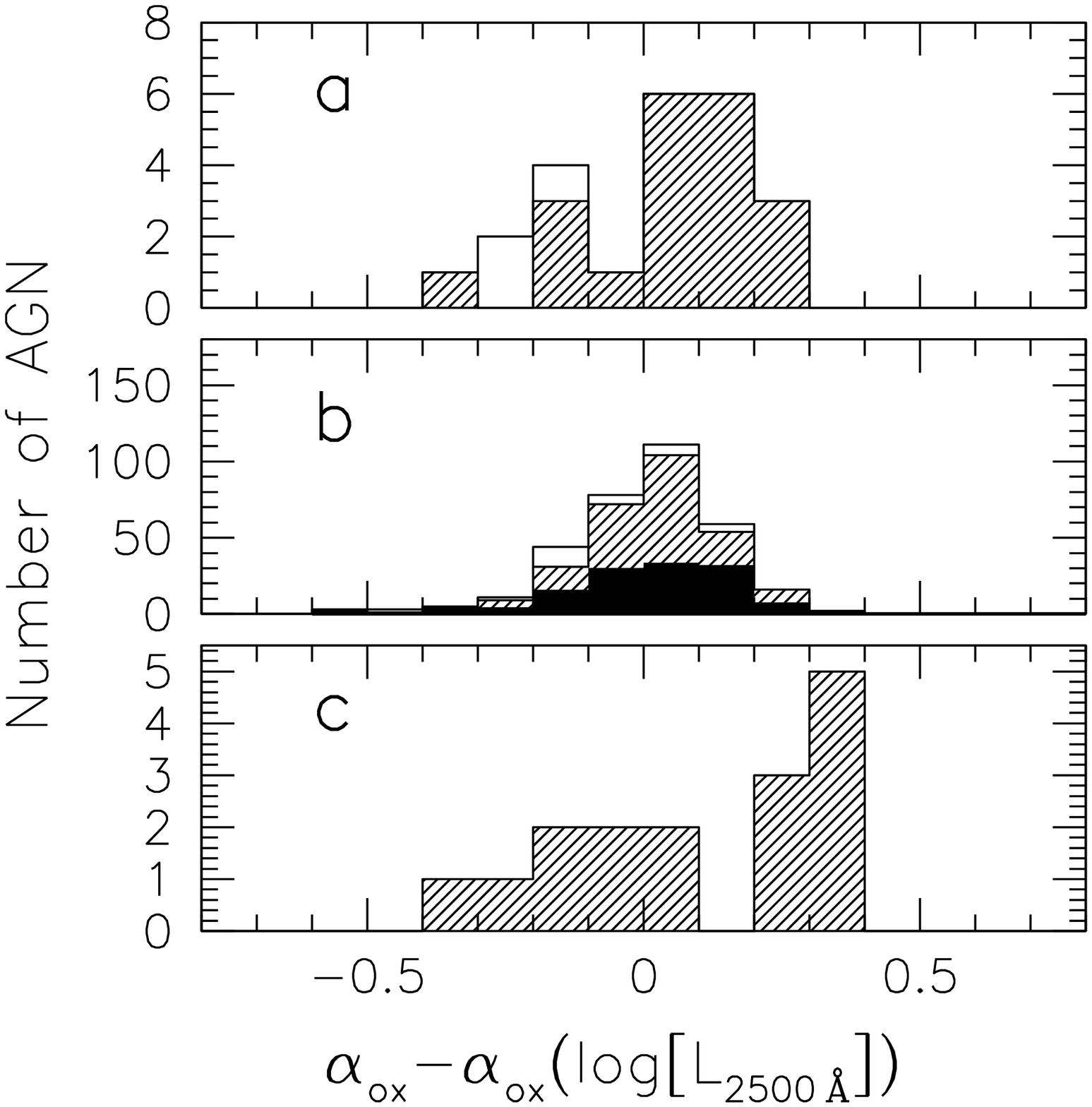}{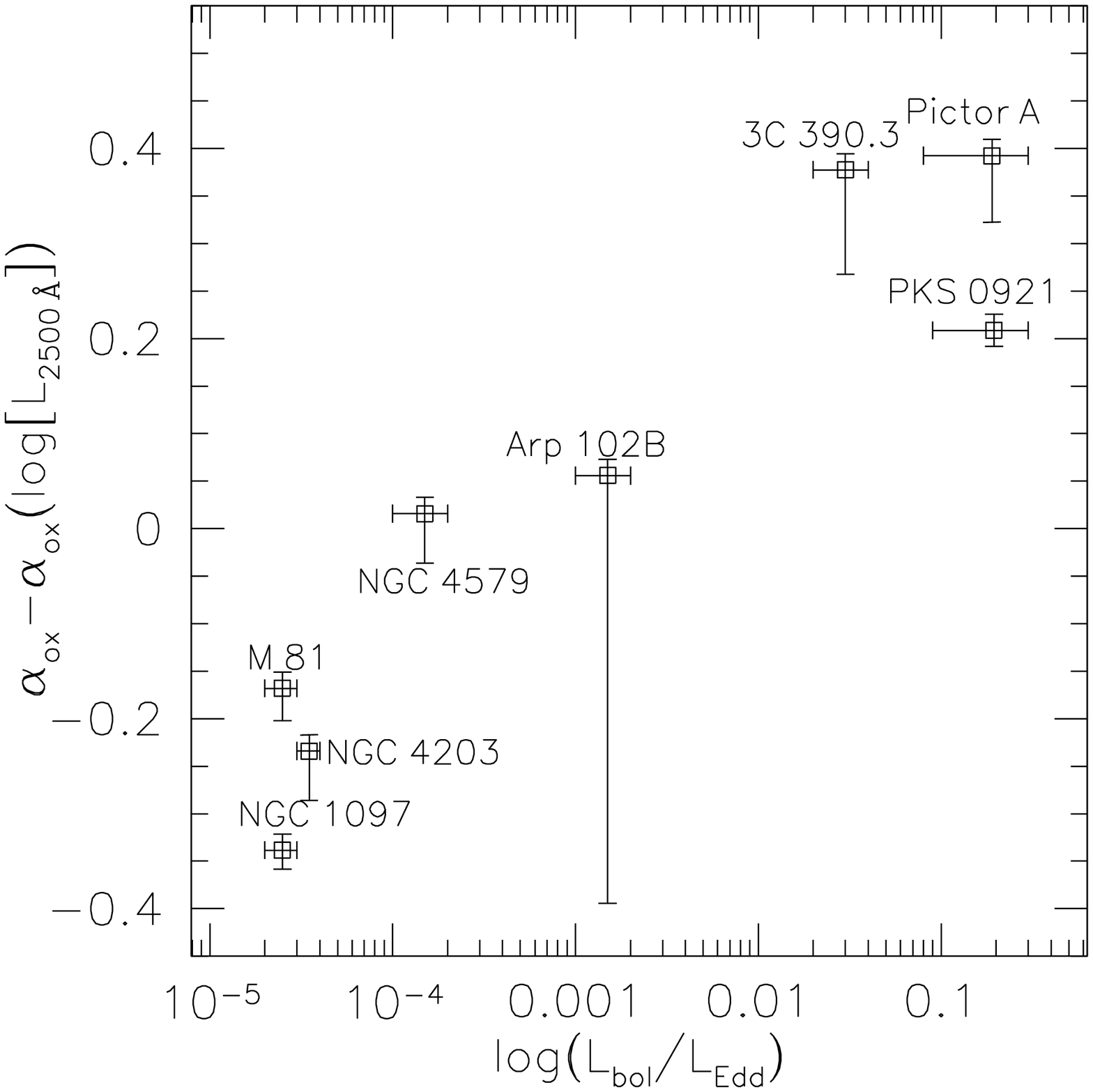}
\caption{\emph{Left:} Comparison between the
  $\alpha\si{ox}-\alpha\si{ox}(l\si{2500\,\AA})$ residual
  distributions for the RQ AGN from the main sample of double-peaked
  emitters (panel a, hatched), the \protect \citet{Steffen05} sample
  (panel b, hatched), the luminosity-matched subsample of
  \citet{Steffen05} (panel b, solid), and the RL double-peaked
  emitters from both the main and auxiliary samples (panel c,
  hatched). The open histograms in each panel indicate the
  $\alpha\si{ox}-\alpha\si{ox}(l\si{2500\,\AA})$ limits. \emph{Right:}
  $\alpha\si{ox}-\alpha\si{ox}(l\si{2500\,\AA})$ residuals vs.
  bolometric luminosity as a fraction of the Eddington luminosity for
  eight auxiliary sample RL double-peaked emitters. This plot includes
  objects whose black hole masses were obtained from stellar velocity
  dispersion measurements.
 \label{aox2}}
\end{figure*}

The top-left panel of Figure~\ref{aox2} presents the histogram of
$\alpha\si{ox}$ residuals for the RQ main sample of double-peaked
emitters, obtained by subtracting the expected
$\alpha\si{ox}(l\si{2500\,\AA})$ from each observed $\alpha\si{ox}$,
in comparison with that of the full sample and a luminosity-matched
subsample ($l\si{2500\,\AA}<30$) from \citet{Steffen05}, shown in the
middle left panel of Figure~\ref{aox2}. The comparison with the
luminosity-matched subsample is useful, as there are some indications
that the $\alpha\si{ox}$-$l\si{2500\,\AA}$ relation might be
non-linear \citep{Steffen05}. The Kaplan-Meier (K-M) estimator mean
value of the $\alpha\si{ox}-\alpha\si{ox}(l\si{2500\,\AA})$ residuals
for the RQ main sample is
$\left<\alpha\si{ox}-\alpha\si{ox}(l\si{2500\,\AA})\right>=0.01\pm0.04$,
consistent with zero, as is the equivalent K-M estimate,
$\left<\alpha\si{ox}-\alpha\si{ox}(l\si{2500\,\AA})\right>=-0.02\pm0.02$,
for the luminosity-matched \citet{Steffen05} subsample. Both the Gehan
and the logrank tests ($T=1.0$, $P=32$\% and $T=0.4$, $P=72$\%)
confirm that there is no significant difference in the cumulative
distributions of the RQ subsample of the main sample and the
luminosity-matched \citet{Steffen05} subsample. The conclusions remain
unchanged if we include the full \citet{Steffen05} sample. Both
$\alpha\si{ox}$ residual distributions are consistent with being
Gaussian, with a width of $0.17\pm0.04$ for the RQ main sample and
$0.14\pm0.01$ for the luminosity-matched \citet{Steffen05} subsample
and small positive means ($0.07\pm0.04$ and $0.04\pm0.01$,
respectively).  The specific Gaussian parameters (especially the
distribution means) are slightly dependent on the size of the bins
used and the fact that we ignore a small number of $\alpha\si{ox}$
limits when performing the Gaussian fits, but the fit parameters are
generally within the 1$\sigma$ errors quoted above. In all cases the
RQ double-peaked emitters tend to have a broader $\alpha\si{ox}$
residual distribution than the corresponding luminosity-matched
\citet{Steffen05} subsample distribution, but the difference is at the
1$\sigma$ level. If this result is confirmed in larger samples, it
could indicate larger X-ray and/or UV variability for the
double-peaked emitters. Our current results indicate that the
double-peak emitters as a class cannot have dramatically different
variability properties than normal active galaxies of similar
optical/UV luminosity.

The left panel of Figure~\ref{aox2} shows that RL double-peaked
emitters have a significantly different $\alpha\si{ox}$ distribution
than RQ double-peaked emitters and the \citet{Steffen05} (full and
luminosity-matched) samples. Because of the difference in censoring
patterns between the \citet{Steffen05} sample and the sample of RL
double-peaked emitters, the Gehan and logrank tests, which assume no
such difference exists, were not used in this comparison. A Peto \&
Prentice test confirms that the RL double-peaked emitters have
$\alpha\si{ox}$ residuals which are substantially different from those
of normal RQ AGN -- $T=2.3$, $P=2$\%. The stronger X-ray emission of
RL double-peaked emitters in comparison to RQ AGN of similar UV
luminosity is also clearly illustrated in Figure~\ref{aox1}, where
eight of the 11 RL double-peaked emitters with $l\si{2500\,\AA}>27.5$
(where the comparison is strictly possible) have $\alpha\si{ox}$
values much flatter than typical RQ AGN. A Peto \& Prentice test
confirms that the $\alpha\si{ox}$ residuals of these 11 RL
double-peaked emitters are significantly different than those of the
RQ AGN in the \citet{Steffen05} sample ($T=4.7$, $P \ll 1$\%).  The
average $\alpha\si{ox}$ residual for these 11 RL double-peaked
emitters is 0.21; the median $\alpha\si{ox}$ residual is 0.27.
Consequently, RL double-peaked emitters are on average $\sim 4
\times$ more X-ray luminous than their RQ counterparts, similar to the
factor $\sim 3$ observed for RL AGN as a whole
\citep[e.g.,][]{XrayRadio}.

\begin{figure*}
\epsscale{1.0}                                       
\plotone{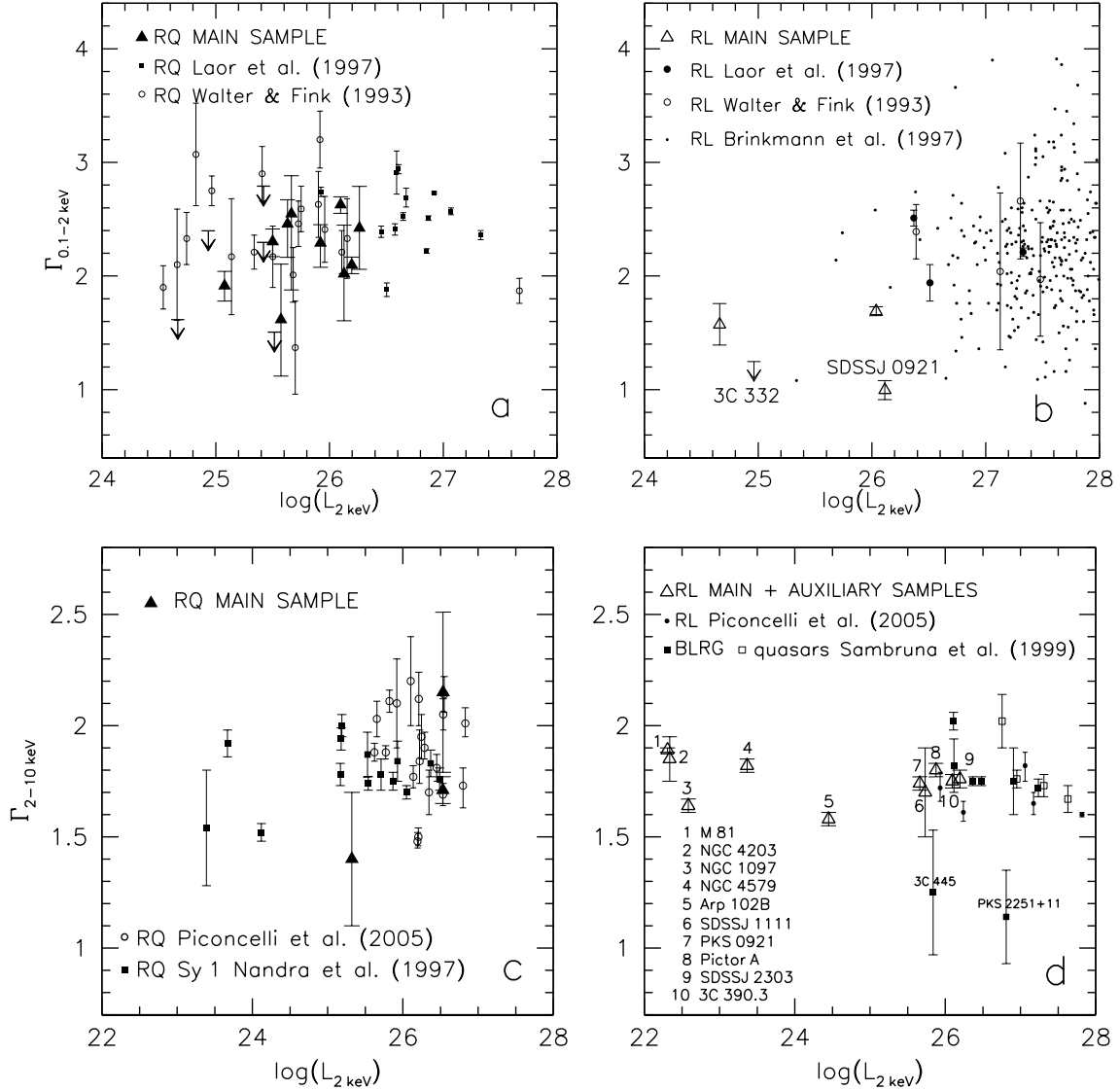}
\caption{The 0.1--2\,keV (panels a and b) and 2--10\,keV (panels c and
d) power-law slopes of RQ (a and c) and RL (b and d) double-peaked
emitters in comparison with other broad-line AGN. The
\hbox{0.1--2\,keV} power-law slopes were obtained using the standard
\emph{ROSAT} hardness ratio HR1, and the \hbox{2--10\,keV} power-law
slopes were obtained through direct spectral fits. Arrows indicate
upper limits on $\Gamma$ (i.e., these AGN were not detected in the
soft [0.1--0.41\,keV] band). The two objects in the lower-right corner
of panel d (PKS\,2251+11 and 3C\,445) are known to have
large intrinsic absorbing column densities of $>10^{22}$\,cm$^{-2}$.
\label{gamma1}} 
\end{figure*}

It is instructive to compare the positions of the higher and lower
luminosity RL double-peaked emitters relative to the extrapolation of
the \citet{Steffen05} relation from Figure~\ref{aox1}.  The higher
luminosity RL objects tend to have flatter $\alpha\si{ox}$ values than
expected for RQ objects with similar luminosities, while the LINER
galaxies appear to have steeper $\alpha\si{ox}$ values. This
difference causes the apparently bimodal $\alpha\si{ox}$-residual
distribution for RL double-peaked emitters shown in in the bottom-left
panel of Figure~\ref{aox2}.  \citet{2peakMbh} have obtained accurate
black-hole mass measurements and estimated the bolometric luminosities
of eight of the RL double-peaked emitters included in our study. The
right panel of Figure~\ref{aox2} shows the $\alpha\si{ox}$ residuals
vs. the bolometric luminosities of these eight sources as a fraction
of the Eddington luminosity. The error bars in the $\alpha\si{ox}$
residuals include the uncertainty due to the intrinsic UV absorption
correction and a constant 10\% error in both the 2500\,\AA\ and 2\,keV
monochromatic-luminosity measurements. It ignores the possibly large
but unknown uncertainties due to the non-simultaneity of the UV and
X-ray observations. To estimate the uncertainty due to an intrinsic UV
absorption correction from the inferred X-ray absorption (or its upper
limit) we assume the \citet{UVext} extinction law and a Galactic
$N_H$/$A_V$ ratio. The uncertainties due to the UV absorption
corrections are typically less than a factor of 3, except for Arp
102B, where it is a factor of 20.  The correlation between
$\alpha\si{ox}-\alpha\si{ox}(l\si{2500\,\AA})$ and $L/L\si{Edd}$ is
strong (partial Kendall's $\tau_{12,3}=0.71$ or $\tau_{12,3}=0.60$)
but significant only at the 3.0\,$\sigma$ or 3.2\,$\sigma$ level when
controlling for the dependence of both variables on luminosity or
redshift.  This observation is consistent with our expectation of a
different SED for AGN with low-radiative-efficiency accretion modes in
comparison to the standard thin-disk and corona models attributed to
higher radiative-efficiency AGN.

\citet{Wu04} have used the FWHM and optical monochromatic-luminosity
measurements of the S03 and \citet{EH94,EH03} samples to obtain rough
estimates of the black-hole masses and accretion rates as a fraction
of the Eddington luminosity. Even though the specific measurements of
\citet{Wu04} are very uncertain (Lewis \& Eracleous 2006 find that two
of the four black-hole masses in common with their sample were
overestimated by an order of magnitude), their general conclusions
that double-peaked emitters have diverse accretion rates were
confirmed by \citet{2peakMbh}. It is therefore encouraging that
\citet{Wu04} find a similar break (at $L/L\si{Edd} \sim 0.001$) in the
$\alpha\si{ox}$ vs.  $L/L\si{Edd}$ relation (see their Figure~4) for
all known RL double-peaked emitters.  Considering the small number of
objects \citep[eight with accurate black-hole mass measurements and a
total of 26 in][] {Wu04} and the large uncertainties in
$\alpha\si{ox}$ (caused by both variability and measurement errors)
for low-luminosity AGN, as well as the strong third-variable
dependences, these findings need confirmation in larger samples.

\subsection{X-ray Spectral Shapes}
\label{Sgamma}

Figure~\ref{gamma1} shows a comparison between the X-ray spectral
slopes measured in the soft (\hbox{0.1--2\,keV}, panels a and b) and
hard (\hbox{2--10\,keV}, panels c and d\footnote{As indicated in
  Table~\ref{tab6}, the hard-band fitting region was extended for five
  of the main-sample double-peaked emitters to include the soft band
  in order to increase the total photon counts to better model the
  intrinsic absorption.}) bands for the double-peaked emitters with
those for similar broad-line AGN which show no evidence of disk
emission in the optical.  The comparison samples of normal AGN were
selected from Laor et al. (1997;
FWHM\si{H$\beta$}$>2000$\,km\,s$^{-1}$), Walter \& Fink (1993,
FWHM\si{H$\beta$}$>2000$\,km\,s$^{-1}$ and $\chi^2/{DoF}<1.1$),
Brinkmann et al. (1997), Piconcelli et al. (2005;
FWHM\si{H$\beta$}$>2000$\,km\,s$^{-1}$), the Seyfert 1 galaxies from
Nandra et al. (1997), and the BLRG and RL quasars from
\citet{Sambruna99}.  Only broad-line AGN
(FWHM\si{H$\beta$}$>2000$\,km\,s$^{-1}$) were selected for this
comparison, since some Narrow-Line Seyfert 1s are known to have
exceptionally steep X-ray spectral slopes \citep[e.g.,][]{NLSy1}, and
their inclusion could bias the results. From Figure~\ref{gamma1}, the
X-ray spectral shapes of double-peaked emitters appear similar to
those of normal AGN, except for the four RL double-peaked emitters in
panel b. A one-dimensional K-S test confirms this statement for the
hard-band power-law slopes shown in Figure~\ref{gamma1}d ($D=0.18$,
$P=98$\%).  There are only three hard-band photon index measurements
for RQ double-peaked emitters in Figure~\ref{gamma1}c; consequently
the K-S test results ($D=0.5$, $P=26$\%) might not be robust.  A
simple Student's $t$-test confirms that the means of the hard-band
photon index distributions for the RQ double-peaked emitters and
normal AGN are consistent, with $t=0.7$ and $P=51$\%.

The presence of photon-index upper limits in panels a and b of
Figure~\ref{gamma1} requires the use of the logrank, Gehan, or Peto \&
Prentice comparison tests. Since the comparison samples do not have
censored data, while the double-peaked emitters do, the Peto \&
Prentice test should give the cleanest results, as it is less
vulnerable to different censoring distributions than the logrank or
Gehan tests \citep{FN85}. According to a Peto \& Prentice test, the
four RL double-peaked emitters in Figure~\ref{gamma1}b, with a K-M
estimator mean of $\left<\Gamma\right>=1.3$ and a dispersion of
$\sigma=0.2$ (compared to $\left<\Gamma\right>=2.22$ and $\sigma=0.04$
for the normal AGN), are significantly different from the normal AGN
($T=4.8$, $P=0$\%). As in the case for 3C\,332 (see discussion below),
this could be an indication of the presence of intrinsic absorption.
Alternatively, it may signify weaker soft excess, which is expected
for the lowest Eddington-ratio AGN (which might lack optically thick
emission from the innermost parts of the accretion disk). According to
a Peto \& Prentice test, the soft-band photon indices of the 15
double-peaked emitters in Figure~\ref{gamma1}a are significantly
different from the comparison AGN ($T=2.4$, $P=2$\%).  The evidence
for this difference is however very weak, relying on the two
double-peaked emitters with the flattest values of $\Gamma$, both of
which are limits. If these two double-peaked emitters are removed from
the comparison, the Peto \& Prentice test gives a 8\% probability that
the remaining double-peaked emitters are indistinguishable from the
normal AGN.

The estimates of the spectral shapes in the \emph{ROSAT} band, which
are based on hardness ratios, depend sensitively on the assumption
that the double-peaked emitters have no intrinsic absorption above the
Galactic values. As reported in \S~\ref{spGamma}, we saw no evidence
of intrinsic absorption in four of the seven main-sample double-peaked
emitters. The remaining three cases (SDSSJ\,918$+$5139,
SDSSJ\,0938$+$0057, and SDSSJ\,1111$+$4820 ) show good evidence of
intrinsic absorption, but the $N\si{H,intr}$ estimates are not robust,
owing to the small number of counts available and the degeneracy
between intrinsic absorption and spectral index in these cases. The
prototype double-peaked emitter Arp\,102B, has a confirmed intrinsic
absorbing column, equivalent to
$N\si{H,intr}=(2.8\pm0.3)\times10^{21}$\,cm$^{-2}$
\citep{Arp102BXray}. In fact, significant intrinsic neutral absorbing
columns are a general characteristic of nearby BLRG \citep[see
Figure~6 of ][]{Sambruna99}. It is therefore plausible that the very
hard power-law slope estimates obtained for some double-peaked
emitters from the observed hardness ratios are a result of intrinsic
absorption that was not taken into account. For example,
SDSSJ\,1617+3222 (3C\,332), one of the six double-peaked emitters with
power-law slope upper limits, has HR1$>0.6$, which, for a Galactic
obscuration of $N_H=2\times10^{20}\,\textrm{cm}^{-2}$ corresponds to
$\Gamma\lesssim1.2$. The 3C\,332 \emph{ROSAT} observation was
previously studied by \citet{3C332xray}, who conclude that the hard
spectral slope obtained by assuming Galactic absorption is probably an
indication of an intrinsic absorber.  Assuming $\Gamma=1.8$,
\citet{3C332xray} derive an obscuring column of $N\si{H,intr}
\approx1\times10^{21}\,\textrm{cm}^{-2}$. The UV-to-X-ray
spectral-index of 3C\,332 is $\alpha\si{ox}=-1.53$, while the value
expected for a RL AGN with comparable UV monochromatic luminosity is
$\approx-1.15$\footnote{Assuming that RL AGN are on the average three
  times brighter at 2\,keV than RQ AGN with comparable UV
  monochromatic luminosity.}, indicative of weaker X-ray emission than
expected and suggestive of intrinsic absorption.

There are a total of four double-peaked emitters in panels a and b of
Figure~\ref{gamma1} with lower $\Gamma\si{0.1--2\,keV}$ than expected,
two RL AGN (3C\,332 and SDSSJ\,0921+4538) with
$\Gamma\si{0.1--2\,keV}<1.5$, and two RQ AGN (SDSSJ\,1101+5122 and
SDSSJ\,1150+0208) with $\Gamma\si{0.1--2\,keV}$ upper limits
($\lesssim1.6$). The two RQ double-peaked emitters with low
$\Gamma\si{0.1--2\,keV}$ upper limits have $\alpha\si{ox}$ values
consistent within 0.05 with the \citet{Steffen05}
\hbox{$\alpha\si{ox}$-$l\si{2500\,\AA}$} relation (indicating no
intrinsic absorption), while SDSSJ\,0921+4538 has
$\alpha\si{ox}=-0.9$, with an expected value of $\alpha\si{ox}=-1.08$
(indicating $\sim 3 \times$ stronger X-ray emission than expected and
no intrinsic absorption).  Taking into account the fact that three of
the four double-peaked emitters with lower than expected
$\Gamma\si{0.1--2\,keV}$ upper limits have small numbers of counts and
only \hbox{0.5--2\,keV}-band detections, and that the observed scatter
in the \hbox{$\alpha\si{ox}$-$l\si{2500\,\AA}$} relation
\citep[$\sim0.11$,][]{aox05}, we consider the above results, with the
exception of those for 3C\,332, to be inconclusive. Longer
observations are necessary to measure accurately the power-law slopes
and intrinsic absorption columns for double-peaked emitters with
unusually low spectral index estimates in the 0.1--2\,keV band.

\section{ENERGY BUDGET}
\label{ebudget}

Using the inner ($\xi_1$) and outer ($\xi_2$) radii of the H$\alpha$
emission region we can estimate the amount of energy available locally
in the disk due to viscous stresses and compare it to the luminosity of
the H$\alpha$ line. An H$\alpha$-line luminosity of 10--20\% or more
of the locally available energy suggests the need for an external
source of illumination.  Assuming a standard \citet{SS73} disk,
\citet{CHF89} and \citet{EH94} estimate the gravitational power output
of the line emitting disk annulus:

\begin{eqnarray}
W\si{disk}(\xi_1,\xi_2)= 7.7\times10^{43}L_{X,42}\times  \nonumber \\  
\left[\frac{1}{\xi_1}\left(1-\sqrt{\frac{8}{3\xi_1}}\right)-
\frac{1}{\xi_2}\left(1-\sqrt{\frac{8}{3\xi_2}}\right)\right]\textrm{erg\,s}^{-1} 
\label{eqn2}
\end{eqnarray}

\noindent
where $L\si{X,42}$ is the \hbox{0.5--2\,keV} \hbox{X-ray} luminosity
in units of 10$^{42}$\,erg\,s$^{-1}$, and we assume for the bolometric
luminosity, $L\si{bol}$, $L\si{X}/L\si{bol}\approx0.1$ and, for an
accretion rate $\dot M$, $L\si{bol}=\zeta \dot M c^2$, with the
efficiency for conversion of energy into radiation $\zeta\approx0.1$.
The H$\alpha$-line luminosities together with $L\si{X,42}$ are listed
in Table~\ref{tab7}. Since only six objects admit axisymmetric-disk
profile fits, and only these allow unique estimates of the emitting
region, we have no direct measurements of $\xi_1$ and $\xi_2$ for each
AGN in the main sample. We do know that the main sample studied here
is representative of the S03 sample and that most circular or
elliptical double-peaked emitters have inner-emission radii of
$200\,R_G<\xi_1<1000\,R_G$ and outer-emission radii of
$1000\,R_G<\xi_2<10000\,R_G$. SDSSJ\,0942+0900 \citep{Wang05}, is an
exception, with robust inner radius estimates
$\xi_1\approx80-100\,R_G$. Based on both observational (as noted
above) and theoretical (the temperature of the disk increases for
small $\xi$ and will likely ionize all hydrogen, preventing
H$\alpha$-line emission) arguments, it is safe to state that
$\xi_1>80\,R_G$ for all known Balmer-line double-peaked emitters. For
the average double-peaked emitter studied by S03, Strateva et al.
(2006, in preparation) and \citet{EH03} find $\xi_1\approx450\,R_G$
and $\xi_2\approx3000\,R_G$.

\begin{figure}
\epsscale{1.0}
\plotone{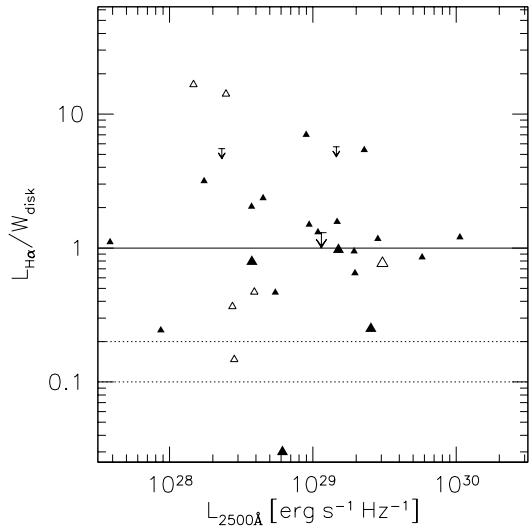}
\caption{Rest-frame UV monochromatic luminosity vs. the ratio of the
  H$\alpha$ line luminosity ($L\si{H$\alpha$}$) and the power
  available locally in the accretion disk ($W\si{disk}$).  Open
  (filled) symbols indicate RL (RQ) AGN, arrows indicate the three
  upper limits. The larger symbols of each type denote the six AGN with
  direct estimates of the H$\alpha$-emission region and $W\si{disk}$ 
  from circular disk fits; the smaller symbols indicate $W\si{disk}$ 
  estimates which assume typical inner and outer emission radii, 
  $\xi_1\approx450\,R_G$ and $\xi_2\approx3000\,R_G$. The dotted 
  lines denote $L\si{H$\alpha$}/W\si{disk}=10$\% and 20\%, while 
  the solid line indicates $L\si{H$\alpha$}=W\si{disk}$.  Note that 
  in 16 of the 29 cases $L\si{H$\alpha$}>W\si{disk}$, i.e. local viscous 
  dissipation in the disk cannot power the H$\alpha$ line emission even 
  in principle.
  \label{diskPower}}
\end{figure}

Using the general sample constraints on $\xi_1$ and $\xi_2$ given
above, we can now estimate $W\si{disk}$ from eqn.~\ref{eqn2} for the
23 double-peaked emitters without axisymmetric-disk fits.
Figure~\ref{diskPower} shows the results assuming a typical
double-peaked emitter with $\xi_1\approx450\,R_G$ and
$\xi_2\approx3000\,R_G$ for each of the 23 objects (i.e.,
$W\si{disk}\approx0.12\times10^{42}L_{X,42}$\,erg\,s$^{-1}$). Only two
of the 29 ($\sim7$\%) double-peaked emitters would generate enough
power locally to produce H$\alpha$ lines of the observed strength
(assuming that as much as 20\% of the power were radiated in the
H$\alpha$ line).  Under the above assumptions for the extent of the
emission region, the H$\alpha$ line luminosity in fact exceeds the
power produced locally in the disk in 16 of the 29 cases.  Even if all
double-peaked emitters without axisymmetric-disk fits had
$\xi\approx80\,R_G$ (and $\xi_2\gtrsim1000\,R_G$),
$W\si{disk}\lesssim0.77\times10^{42}L\si{X,42}$\,erg\,s$^{-1}$; i.e.,
at most 13 out of the 29 ($\sim45$\%) have enough energy generated
locally to produce H$\alpha$ lines of the observed strength. We
conclude that the majority of double-peaked emitters require external
illumination of the disk to account for the strength of the observed
H$\alpha$ line emission.

\section{SUMMARY AND CONCLUSIONS}
\label{sec4}

We have studied the X-ray emission properties of 39 AGN with
double-peaked H$\alpha$ lines, serendipitously observed by
\emph{ROSAT}, \emph{XMM-Newton}, and \emph{Chandra}, controlling for
their UV and radio emission. The main sample consists of 29 objects
selected from the SDSS which are representative of the sample of
double-peaked emitters studied by S03. The remaining 10 objects (the
auxiliary sample) are the best-studied double-peaked emitters (eight
low optical-luminosity BLRG and LINER galaxies, and two RL quasars) with
targeted and serendipitous \hbox{X-ray} observations. The total sample
includes 16 RL and 23 RQ double-peaked emitters.

Based on a comparison of the UV-to-X-ray index, $\alpha\si{ox}$, we
find that RQ double-peaked emitters as a class have comparable
\hbox{X-ray} emission levels to those of normal RQ AGN of similar UV
luminosity. RL double-peaked emitters tend to be brighter in the
\hbox{X-ray} band in comparison with similar-luminosity normal RQ AGN,
as observed for RL AGN in general. The \hbox{0.5--10\,keV} spectral
shapes of the double-peaked emitters studied here are consistent with
those of normal broad-line AGN with similar radio-loudness, with the
possible exception of four RL double-peaked emitters with unusually
flat soft-band spectra. Three double-peaked emitters observed above
2\,keV show signs of large intrinsic absorption
($N\si{H,intr}\gtrsim10^{22}\,\textrm{cm}^{-2}$ in two of the three
cases).

The majority of the double-peaked emitters require external
illumination of the accretion disk, as the power generated locally is
insufficient to produce lines of the observed strength. This result is
more general and was used as an argument against the accretion-disk
origin of broad emission lines in AGN, irrespective of their profiles
\citep[e.g.,][]{SF80}. For AGN in which the
Balmer-lines are double-peaked, this external illumination could be
associated with an elevated structure in the inner disk (e.g., an
\hbox{X-ray} emitting corona, jet, or vertically extended torus).
However, the fact that the \hbox{X-ray} emission of double-peaked
emitters as a class does not differ from that of normal AGN with
similar properties, suggests that a peculiarity of the \hbox{X-ray}
emission structure and/or mechanism is not responsible for the
occurrence of double-peaked Balmer lines in AGN. Despite the
importance of \hbox{X-ray} illumination for the majority of
double-peaked emitters, the strength of the \hbox{X-ray} emission
relative to the optical and UV emission cannot be used to predict if
the broad, low-ionization lines are double-peaked. By extension, the
structure of the inner disk is unlikely to be the only (or the
dominant) reason for the observed double-peaked profiles.

Given the fact that double-peaked emitters as a class appear
indistinguishable in their X-ray properties from normal AGN, the best
course for the future might be to study in detail extreme cases of
double-peaked emitters. Our upcoming \emph{Chandra} observations, for
example, include one of the most luminous known double-peaked emitters
-- an AGN with a double-peaked \ion{Mg}{2} line. In addition, we hope
to obtain high-quality hard-band X-ray observations of the broadest
double-peaked H$\alpha$-line emitters, for which the emission region
of the disk is likely to be in the immediate vicinity or even
coincident with the X-ray emitting structure.

\vskip 0.5in \leftline{Acknowledgments}

IVS, WNB and DPS acknowledge the support of NASA LTSA grant NAG5-13035.
DPS acknowledges the support of NSF grant AST03-07582.

Funding for the SDSS and SDSS-II has been provided by the Alfred
P. Sloan Foundation, the Participating Institutions, the National
Science Foundation, the U.S. Department of Energy, the National
Aeronautics and Space Administration, the Japanese Monbukagakusho, the
Max Planck Society, and the Higher Education Funding Council for
England. The SDSS Web Site is http://www.sdss.org/. The SDSS is
managed by the Astrophysical Research Consortium for the Participating
Institutions. The Participating Institutions are the American Museum
of Natural History, Astrophysical Institute Potsdam, University of
Basel, Cambridge University, Case Western Reserve University,
University of Chicago, Drexel University, Fermilab, the Institute for
Advanced Study, the Japan Participation Group, Johns Hopkins
University, the Joint Institute for Nuclear Astrophysics, the Kavli
Institute for Particle Astrophysics and Cosmology, the Korean
Scientist Group, the Chinese Academy of Sciences (LAMOST), Los Alamos
National Laboratory, the Max-Planck-Institute for Astronomy (MPA), the
Max-Planck-Institute for Astrophysics (MPIA), New Mexico State
University, Ohio State University, University of Pittsburgh,
University of Portsmouth, Princeton University, the United States
Naval Observatory, and the University of Washington.

This research has made use of the Tartarus (Version 3.1) database,
created by Paul O'Neill and Kirpal Nandra at Imperial College London,
and Jane Turner at NASA/GSFC. Tartarus is supported by funding from
PPARC, and NASA grants NAG5-7385 and NAG5-7067.  This research has
made use of the NASA/IPAC Extragalactic Database (NED) which is
operated by the Jet Propulsion Laboratory, California Institute of
Technology, under contract with the National Aeronautics and Space
Administration. 

\begin{figure}
\epsscale{1.0}
\plotone{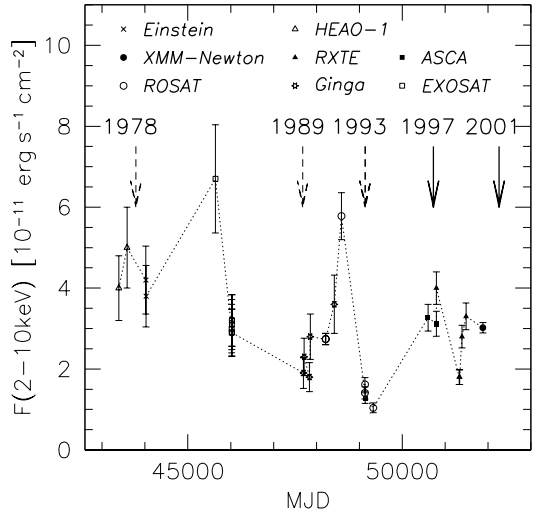}
\caption{The 2--10\,keV band variability of SDSSJ\,2304$-$0841
\citep[MCG-2-58-22, following][]{MCG25822}. The solid arrows indicate
two optical spectra with double-peaked Balmer line profiles (obtained
in 1997 and 2001), the dashed arrows indicate three optical spectra
with single-peaked lines \protect \citep[obtained in 1978, 1989, and
1993;][]{Durret,Marziani,Argote}.
\label{926var}} 
\end{figure}

\appendix
\section{THE X-RAY AND OPTICAL VARIABILITY OF SDSSJ\,2304$-$0841}
\label{sec:app}

SDSSJ\,2304$-$0841 is highly variable on long timescales (6--20
years). \citet{MCG25822} cite evidence of secular changes in flux
accompanied by strong, shorter lived ($\sim$1--2\,years)
flares\footnote{\citet{MCG25822} were interested in the X-ray flux
variability only; they do not discuss the spectral shape and
Fe-K$\alpha$ variability or their relation to the variability in the
optical line profile.}. Figure~\ref{926var} shows the {2--10\,keV}
band variability (following closely\footnote{We include the latest
\emph{XMM-Newton} point, replace the \emph{ROSAT} fluxes from Choi et
al. 2002 with our estimates, and the \emph{ASCA} fluxes with the
latest Tartarus database results obtained from HEASARC.} Figure~1 of
\citet{MCG25822}) with arrows indicating whether the H$\alpha$-line
profiles were single or double-peaked. The 2002 optical SDSS
observation shows complex H$\alpha$ and H$\beta$ line profiles with at
least two peaks and an extended red tail. A 1997 observation by
\citet{Argote} shows simple double-peaked profiles for both Balmer
lines, while a May 1993 observation of H$\beta$ by \citet{Marziani}
and a 1989 observation by \citet{Argote} show simple single-peaked
profiles characteristic of normal AGN. The earliest known optical
spectrum of SDSSJ\,2304$-$0841 was taken in 1978 by \citet{Durret};
judging by the spectrum in that paper the H$\beta$ line was single
peaked. The 1997 observation of a double-peaked H$\alpha$ line
coincides with one of the X-ray flares identified by \citet{MCG25822},
while the three observations of single-peaked H$\alpha$ lines are in
non-flare regions according to \citet{MCG25822}; we have no X-ray data
coincident with the SDSS double-peaked profile (the
\emph{XMM-Newton}/\emph{BeppoSAX} data were taken one year before the
SDSS optical spectrum). From Figure~\ref{926var} it is tempting to
conclude that the occurrence of disk emission in the optical is
unrelated to flares in the X-ray flux. This result, however, is highly
uncertain, considering the sparse sampling time, the different X-ray
bands observed, and the (significant) uncertainties of instrument
cross-calibration. What is obvious from Figure~\ref{926var}, is that
the appearance of optical double-peaked lines in not predominantly
associated with either low- or high-X-ray-flux states. The X-ray
spectral analysis presented in \S~\ref{spec1} further suggests that
the appearance of optical double-peaked lines is not correlated
with the steepness of the hard-band spectrum or the presence of a
strong Fe~K$\alpha$ line.


\end{document}